\newif\ifarxiv

\arxivtrue

\ifarxiv
\documentclass[final,5p,times,twocolumn]{elsarticle}
\else
\documentclass[final,3p,times]{elsarticle}
\fi

\usepackage[utf8]{inputenc}
\usepackage[T1]{fontenc}

\graphicspath{{graphics/}}

\usepackage{amsmath}
\usepackage{amsfonts}
\usepackage{amssymb}
\usepackage{float}

\usepackage[normalem]{ulem}

\usepackage{booktabs}
\usepackage{tabularx}
\usepackage{threeparttable}
\usepackage{siunitx}

\usepackage{url}
\usepackage[colorlinks=true, citecolor=blue, linkcolor=blue, filecolor=blue,urlcolor=blue]{hyperref}

\usepackage[gen]{eurosym}

\def\co{CO${}_2$}

\begin{document}
%\linenumbers
\begin{frontmatter}



  \title{Response to
  `Burden of proof: A comprehensive review of the feasibility of 100\% renewable-electricity systems'}


\author[kit,fias]{T.~W.~Brown\corref{cor1}}
\ead{tom.brown@kit.edu}
\author[csir]{T.~Bischof-Niemz}
\author[delft]{K.~Blok}
\author[lapp]{C.~Breyer}
\author[aalborg]{H.~Lund}
\author[aalborgc]{B.V.~Mathiesen}

\cortext[cor1]{Corresponding author}

\address[kit]{Institute for Automation and Applied Informatics, Karlsruhe Institute of Technology, 76344 Eggenstein-Leopoldshafen,
Germany}
\address[fias]{Frankfurt Institute for Advanced Studies, Ruth-Moufang-Straße 1, 60438 Frankfurt am Main, Germany}
\address[csir]{Energy Centre, Council for Scientific and Industrial Research, Meiring Naude Road, Pretoria, South Africa}
\address[delft]{Delft University of Technology, Chair of Energy Systems Analysis, Faculty Technology, Policy and Management, Jaffalaan 5, 2628 BX Delft, The Netherlands}
\address[lapp]{Lappeenranta University of Technology, School of Energy Systems, Skinnarilankatu 34, 53850 Lappeenranta, Finland}
\address[aalborg]{Department of Development and Planning, Aalborg University, Rendsburggade 14, 9000 Aalborg, Denmark}
\address[aalborgc]{Department of Development and Planning, Aalborg University, A.C. Meyers Vænge 15, 2450 Copenhagen SV, Denmark}

\begin{abstract}

A recent article `Burden of proof: A comprehensive review of the
feasibility of 100\% renewable-electricity systems' claims that many
studies of 100\% renewable electricity systems do not demonstrate
sufficient technical feasibility, according to the criteria of the
article's authors (henceforth `the authors').  Here we analyse the
authors' methodology and find it problematic. The feasibility criteria
chosen by the authors are important, but are also easily addressed at
low economic cost, while not affecting the main conclusions of the
reviewed studies and certainly not affecting their technical
feasibility. A more thorough review reveals that all of the issues
have already been addressed in the engineering and modelling
literature. Nuclear power, which the authors have evaluated positively
elsewhere, faces other, genuine feasibility problems, such as the
finiteness of uranium resources and a reliance on unproven
technologies in the medium- to long-term. Energy systems based on
renewables, on the other hand, are not only feasible, but already
economically viable and decreasing in cost every year.

\end{abstract}


\end{frontmatter}

\setcounter{tocdepth}{2}
\tableofcontents

\section{Introduction}\label{sec:intro}

There is a broad scientific consensus that anthropogenic greenhouse
gas emissions should be rapidly reduced in the coming decades in order
to avoid catastrophic global warming \cite{IPCC2014-synthesis}.  To
reach this goal, many scientific studies
(\cite{Cochran2014,Jacobson2011a,Jacobson2011b,wwf,er2012,Jacobson08122015,ELLISTON2012606,Elliston,LUND2009524,CONNOLLY2011502,PWC2010,budischak2013,GROSSMANN2013831,PLEMANN201422,BOGDANOV2016176,HUBER2015180,Czisch2005,STEINKE2013826,RASMUSSEN2012642,Hagspiel,Connolly20161634,BUSSAR20161,DOMINKOVIC20161517,GROSSMANN2014983,HUBER2015235,BECKER2014443,CHILD2016517,MATHIESEN2015139,Palzer20141019,IJSEPM497,FERNANDES201451,MOELLER201439,PIP:PIP2885,PIP:PIP2950,Jacobson2017,JACOBSON2018,su9020233,en10050583,PLEMANN201719,Schlachtberger2017,Hoersch2017,GILS2017173,ERIKSEN2017913,CEBULLA2017211,Brown2018,PLEMANN2017,10.1371/journal.pone.0173820,en10081171,GULAGI2017,10.1371/journal.pone.0180611,BLAKERS2017471,LU2017663,barbosa2016,Aghahosseini2017,SADIQA2018518,Caldera2018,KILICKAPLAN2017218,CHILD2017410,GILS2017342,CHILD201749} are discussed in this article)
have examined the potential to replace fossil fuel energy sources with
renewable energy. Since wind and solar power dominate the expandable
potentials of renewable energy \cite{Jacobson2011a}, a primary
focus for studies with high shares of renewables is the need to
balance the variability of these energy sources in time and
space against the demand for energy services.

The studies that examine scenarios with very high shares of renewable
energy have attracted a critical response from some quarters,
particularly given that high targets for renewable energy are
now part of government policy in many countries \cite{Martinot2007,REN17}.  Critics
have challenged studies for purportedly not taking sufficient
account of: the variability of wind and solar
\cite{TRAINER20104107,TRAINER2012476}, the scaleability of some
storage technologies \cite{Clack2017}, all aspects of system costs
\cite{TRAINER20104107,TRAINER2012476}, resource constraints
\cite{TRAINER2013845,Loftus2015}, social acceptance constraints
\cite{Loftus2015}, energy consumption beyond the electricity sector
\cite{Loftus2015}, limits to the rate of change of the energy
intensity of the economy \cite{Loftus2015} and limits on capacity
deployment rates \cite{Smil2010,Loftus2015}. Many of these criticisms
have been rebutted either directly
\cite{DELUCCHI2012482,JACOBSON2013641,Jacobson2017res} or are addressed elsewhere in the
literature, as we shall see in the following sections.

In the recent article `Burden of proof: A comprehensive review of the
feasibility of 100\% renewable-electricity systems' \cite{burden} the
authors of the article (henceforth `the authors') analysed 24
published studies (including \cite{Jacobson2011a,Jacobson2011b,wwf,er2012,Jacobson08122015,ELLISTON2012606,Elliston,PWC2010,budischak2013,LUND2009524,CONNOLLY2011502})
of scenarios for highly renewable electricity systems, some regional
and some global in scope. Drawing on the criticisms outlined above,
the authors chose feasibility criteria to assess the studies,
according to which they concluded that many of the studies do not rate
well.

In this response article we argue that the authors' chosen feasibility
criteria may in some cases be important, but that they are all easily
addressed both at a technical level and economically at low cost.  We
therefore conclude that their feasibility criteria are not useful and
do not affect the conclusions of the reviewed studies. Furthermore, we
introduce additional, more relevant feasibility criteria, which
renewable energy scenarios fulfil, but according to which
nuclear power, which the authors have evaluated positively elsewhere
\cite{Brook2014,Heard2017,Brook2018}, fails to demonstrate adequate
feasibility.

In Section \ref{sec:feasvia} we address the definition and relevance
of feasibility versus viability; in Section \ref{sec:criteria} we
review the authors' feasibility criteria and introduce our own additional
criteria; in Section \ref{sec:other} we address other issues raised by
\cite{burden}; finally in Section \ref{sec:conclusions} conclusions
are drawn.

\section{Feasibility versus viability}\label{sec:feasvia}

Early in their methods section, the authors define \emph{feasibility}
to mean that something is technically possible in the world of physics `with
current or near-current technology'. They distinguish feasibility from
socio-economic \emph{viability}, which they define to mean whether it
is possible within environmental and social constraints and at a
reasonable cost. While there is no widely-accepted definition of
feasibility \cite{IPCC2014-III-ATP}, other studies typically include
economic feasibility in their definition
\cite{Kaldellis2010102,Bekele2010,CASTROSANTOS2016868,Rodrigues2016,TSUPARI20171040},
while others also consider social and political constraints
\cite{RIDJAN201376,Loftus2015}. For the purposes of this response
article, we will keep to the authors' definitions of feasibility and
viability.

One reason that few studies focus on such a narrow technical
definition of feasibility is that, as we will show in the
sections below, there are solutions using today's
technology for all the feasibility issues raised by the authors.
The more interesting question, which is where most studies rightly
focus, is how to reach a high share of renewables in the most
cost-effective manner, while respecting environmental, social and
political constraints. In other words, viability is where the real
debate should take place.  For this reason, in this paper we will
assess both the feasibility and the viability of renewables-based
energy systems.

Furthermore, despite their declared focus on feasibility, the authors
frequently mistake viability for feasibility. Examples related to
their feasibility criteria are examined in more detail below, but
even in the discussion of specific model results there is confusion. The
authors frequently quote from cost-optimisation studies that `require'
certain investments. For example they state that \cite{macdonald2017}
`required 100~GWe of nuclear generation and 461~GWe of gas' and
\cite{Rodriguez2013} `require long-distance interconnector capacities
that are 5.7 times larger than current capacities'. Optimisation
models find the most cost-effective (i.e. viable) solutions within
technical constraints (i.e. the feasible space). An optimisation
result is not necessarily the only feasible one; there may be many
other solutions that simply cost more. More analysis is needed to find
out whether an investment decision is `required' for feasibility or
simply the most cost-effective solution of many. For example, the
100~GWe of nuclear in \cite{macdonald2017} is fixed even before the
optimisation, based on existing nuclear facilities, and is therefore not the
result of a feasibility study. However, the authors do
acknowledge that their transmission feasibility criteria `could
arguably be regarded as more a matter of viability than feasibility'.

Finally, when assessing economic viability, it is important to keep a sense of
perspective on costs. If Europe is taken as an example, Europe pays
around 300-400 billion \euro{} for its electricity annually.\footnote{Own calculation based on price and (incomplete) consumption data from Eurostat \cite{eurostat} for 2015. It includes energy supply (around 50\%), network costs (around 20\%), taxes and surcharges (around 30\%); it excludes indirect costs, such as those caused by environmental pollution \cite{millstein2017} and climate change \cite{IPCC2014-economic}.} EU GDP in 2016
was 14.8 trillion \euro{}
\cite{eurostat}. Expected electricity
network expansion costs in Europe of 80 billion \euro{}
until 2030 \cite{TYNDP2016} may sound high, but once  these costs are annualised
(e.g. to 8 billion \euro{}/a), it amounts to only 2\% of total spending on
electricity, or 0.003 \euro{}/kWh.

\section{Feasibility Criteria}\label{sec:criteria}

The authors define feasibility criteria and rate 24 different studies
of 100\% renewable scenarios against these criteria. According to the
chosen criteria, many of the studies do not rate highly.

In the sections below we address each feasibility criterion
mentioned by the authors, and some additional ones which we believe are more pertinent. In
addition, we discuss the socio-economic viability of the feasible solutions.

We observe that the authors' choice of criteria, the weighting given to
them and some of the scoring against the criteria are somewhat
arbitrary. As argued below, there are other criteria that the authors
did not use in their rating that have a stronger impact on feasibility
(such as resource constraints and technological maturity); based on
the literature review below, the authors' criteria would receive a
much lower weighting than these other, more important criteria; and the scoring of some of the criteria,
particularly for primary energy, transmission and ancillary services,
seems coarse and subjective. Regarding the scoring, for demand
projections the studies are compared with a spectrum from the mainstream
literature, but no uncertainty bound is given, just a binary score;
for transmission there is no nuance between studies that use blanket
costs for transmission, or only consider cross-border capacity, or
distribution as well as transmission networks; and no weighting is
given to the importance of the different ancillary services.

Finally, note that while some of the studies chosen by the authors
consider the electricity sector only, other studies include energy
demand from other sectors such as transport, heating and industry,
thereby hindering comparability between the studies.

\subsection{Their Feasibility Criterion 1: Demand projections}
\label{sec:demand}

The authors criticise some of the studies for not using plausible
projections for future electricity and total energy
demand. In particular, they claim that reducing global primary energy
consumption demand is not consistent with projected population growth
and development goals in countries where energy demand is currently
low.

Nobody would disagree with the authors that any future energy scenario
should be compatible with the energy needs of every citizen of the
planet. A reduction in electricity demand, particularly if heating,
transport and industrial demand is electrified, is also unlikely to be
credible. For example, both the Greenpeace Energy
[R]evolution \cite{er2012,er2015} and WWF \cite{wwf} scenarios,
criticised in the paper, see a significant increase in global
electricity consumption; another recent study \cite{PIP:PIP2950} of
100\% renewable electricity for the globe foresees a doubling of electricity
demand between 2015 and 2050, in line with IEA estimates for electricity \cite{weo2016}.

However, the authors chose to focus on primary energy, for which the situation is more
complicated, and it is certainly plausible to decouple primary energy
consumption growth from meeting the planet's energy needs. Many
countries have already decoupled primary energy supply from economic
growth; Denmark has 30 years of proven history in reducing the energy
intensity of its economy \cite{Lund2014}.

There are at least three points here: i) primary energy consumption
automatically goes down when switching from fossil fuels to wind,
solar and hydroelectricity, because they have no conversion losses
according to the usual definition of primary energy; ii) living
standards can be maintained while increasing energy efficiency; iii)
renewables-based systems avoid the significant energy usage of mining,
transporting and refining fossil fuels and uranium.

\begin{figure}[t]
  \centering
      %trim={<left> <lower> <right> <upper>}
\includegraphics[width=\linewidth,trim=2.7cm 20cm 3.2cm 1.8cm,clip=true]{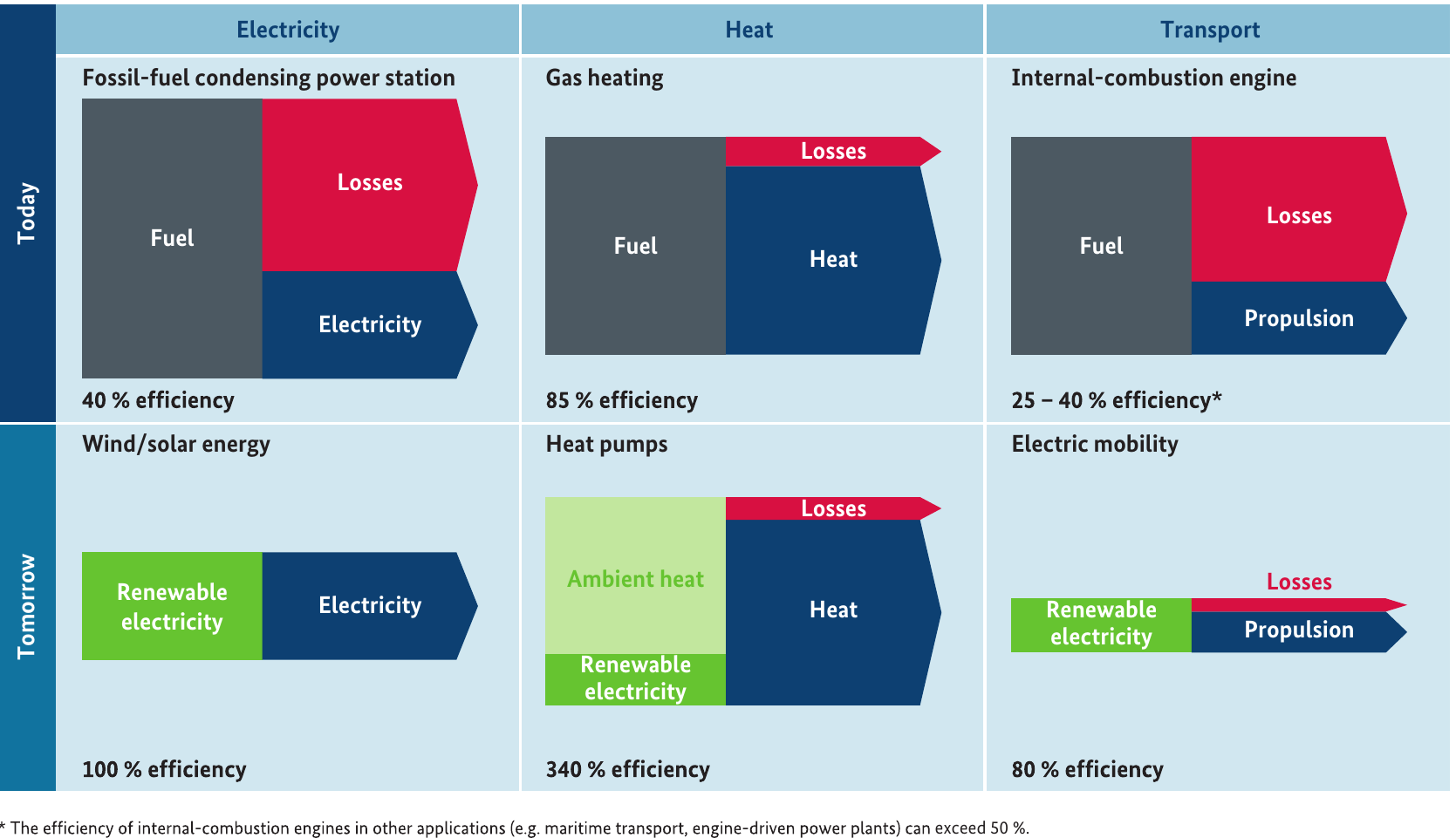}
\caption{\label{fig:weissbuch} Primary energy consumption (grey and green) versus useful energy services (blue) in today's versus tomorrow's energy system. (Reproduced with permission from \cite{whitepaper2015}, page 86; based on \cite{iwesgreen2015})}
\end{figure}

Figure \ref{fig:weissbuch} illustrates how primary energy consumption can decrease
by switching to renewable energy sources, with no change in the energy services (blue) delivered.
Using the `physical energy accounting method' used by the IEA, OECD, Eurostat
and others, or the `direct equivalent method' used by the IPCC, the
primary energy consumption of fossil fuel power plants corresponds to
the heating value, while for wind, solar and hydro the electricity output
is counted. This automatically leads to a reduction in the primary
energy consumption of the electricity sector when switching to wind,
solar and hydro, because they have no conversion losses (by this
definition).

In the heating sector, fossil-fuelled boilers dominate today's heating
provision; here, primary energy again corresponds to the heating value
of the fuels.  For heat pumps, the heat taken from the
environment is sometimes counted as primary energy
\cite{eurostatRE,WB2014}, sometimes not \cite{wwf}; in the latter case
the reduction in primary energy consumption is 60-75\%
\cite{Staffell2012}, depending on the location and technology, if
wind, solar and hydro power are used. Cogeneration of heat and power will also reduce primary
energy consumption. In addition, district heating can be
used to recycle low-temperature heat that would otherwise be lost,
such as surplus heat from industrial processes
\cite{MATHIESEN2012160,LUND20141,CONNOLLY2014475}.  For biomass, solar
thermal heating and resistive electric heating from renewables there
is no significant reduction in primary energy compared to
fossil-fuelled boilers.

In transport, the energy losses in an internal combustion engine mean
that switching to more efficient electric vehicles running on
electricity from wind, solar and hydro will reduce primary energy
consumption by 70\% or more \cite{Brown2018} for the same service.

If statistics from the European Union in 2015 \cite{eurostat-eb} are
taken as an example, taking the steps outlined in Figure
\ref{fig:weissbuch} would reduce total primary energy consumption by
49\%\footnote{This would involve switching from thermal power plants
  (average efficiency 35\% \cite{eurostat-eb}) to wind and solar
  generators for all electricity, using heat pumps (average
  coefficient of performance of 3) for space and water heating, and
  using electricity instead of internal combustion in road vehicles
  (reducing final energy consumption here by a factor of 3.5
  \cite{Brown2018}). No reduction in primary energy is assumed for
  remaining energy sectors (non-electric industrial demand, aviation
  and shipping).}  without any change in the delivered energy
services. (Final energy consumption would also drop by 33\%.)
% See energy/structure.org for full calculation
A reduction of total primary energy of 49\% would allow a near
doubling of energy service provision before primary energy consumption
started to increase. This is even before efficiency measures and the
consumption from fuel processing are taken into account.

The primary energy accounting of different energy sources presented in
this example is already enough to explain the discrepancies between
the scenarios plotted in Figure 1 of \cite{burden}, where the median
of non-NGO global primary energy consumption increases by around 50\%
between 2015 and 2050, while the NGOs Greenpeace and WWF see light
reductions.
%>>> 820/560  1.4642857142857142
As an example of a non-NGO projection with high primary energy demand, many IPCC scenarios with reduced greenhouse gas emissions
rely on bioenergy, nuclear and carbon capture from combustion
\cite{vanVuuren2011}, whereas the NGOs Greenpeace \cite{er2012} and WWF
\cite{wwf} have high shares of wind and solar. The IPCC scenarios see
less investment in wind and solar because of conservative
cost assumptions, with some assumptions for solar PV that are 2-4
times below current projections \cite{Schellnhuber2016,PIP:PIP2885};
with improved assumptions, some authors calculate that PV could
dominate global electricity by 2050 with a share of 30–50\%
\cite{Creutzig2017}. Another study of 100\% renewable energy across all energy sectors in Europe \cite{Connolly20161634} sees a 10\% drop in primary energy supply compared to a business-as-usual scenario for 2050, with bigger reductions if synthetic fuels for industry are excluded.

The authors chose to concentrate on primary energy consumption, but
for renewables, as argued above, it can be a misleading metric (see
also the discussion in \cite{WB2014}). The definitions of both primary
and final energy are suited for a world based on fossil fuels. What
really matters is meeting people's energy needs (the blue boxes in
Figure \ref{fig:weissbuch}) while also reducing greenhouse gas
emissions.

Next we address energy efficiency that goes beyond just switching fuel
source. There is plenty of scope to maintain living standards while
reducing energy consumption: improved building insulation and design
to reduce heating and cooling demand, more efficient electronic
devices, efficient processes in industry, better urban design to lower
transport demand, more public transport and reductions in the
highest-emission behaviour. These efficiency measures are feasible,
but it is not clear that they will all be socio-economically viable.

For example, in a study for a 100\% renewable German energy system
(including heating and transport) \cite{Palzer20141019} scenarios were considered
where space heating demand is reduced by between 30\% and 60\%
using different retrofitting measures. Another study for cost-optimal 100\% renewables in Germany \cite{IEESWV} shows similar reductions in primary energy in the heating
sector from efficiency measures and the uptake of cogeneration and heat pumps.

The third point concerns the upstream costs of conventional fuels. It
was recently estimated that 12.6\% of all end-use energy worldwide is
used to mine, transport and refine fossil fuels and uranium \cite{Jacobson2017};
renewable scenarios avoid this fuel-related consumption.

One final, critical point: even if future demand is higher than
expected, this does not mean that 100\% renewable scenarios are
infeasible.  As discussed in Section \ref{sec:fuel}, the global
potential for renewable generation is several factors higher than any
demand forecasts. There is plenty of room for error if forecasts prove to underestimate demand growth: an investigation into the United States
Energy Information Administration's Annual Outlook \cite{EIAstudy2008} showed systematic underestimation of total energy demand by an average of 2\% per year after controlling for other sources of projection errors; over 35 years this would lead to an underestimate of around factor 2 (assuming other sources of growth are not excessive); reasonable global potentials for renewable energy could generate on average around 620~TW \cite{Jacobson2011a}, which is a factor 30 higher than business-as-usual forecasts for average global end-use energy demand of 21~TW in 2050 \cite{Jacobson2017}.

%>>> 622/21
%29.61904761904762

% 1.02**35 = 2

%example of addition to 1.05 other sources of growth:

%>>> 1.02**35
%1.9998895526624565
%>>> 1.05**35
%5.5160153675922565
%>>> 1.07**35
%10.676581484615452

\subsection{Their Feasibility Criterion 2a: Simulation time resolution}
\label{sec:resolution}

The authors stress that it is important to model in a high time
resolution so that all the variability of demand and renewables is
accounted for. They give one point to models with hourly resolution
and three points to models that simulate down to 5 minute
intervals.

It is of course important that models have enough time resolution to
capture variations in energy demand (e.g. lower electricity
consumption at 3am than at 3pm) and variations in wind and solar
generation, so that balancing needs, networks and other flexibility
options can be dimensioned correctly. However, the time resolution
depends on the area under consideration, since short-term weather
fluctuations are not correlated over large distances and therefore
balance out. This criterion should rather read `the time resolution
should be appropriate to the size of the area being studied, the
weather conditions found there and the research question'.
Models for whole countries typically use hourly simulations, and we will argue
that this is sufficient for long-term energy system planning.

After all, why do the authors stop at 5 minute intervals? For a single
wind turbine, a gust of wind could change the feed-in within seconds
(the inertia of the rotor stops faster changes). Similarly, a cloud
could cover a small solar panel in under a second. Individuals can
change their electricity consumption at the flick of a switch.

\begin{figure}[t]
\centering
\includegraphics[width=\linewidth]{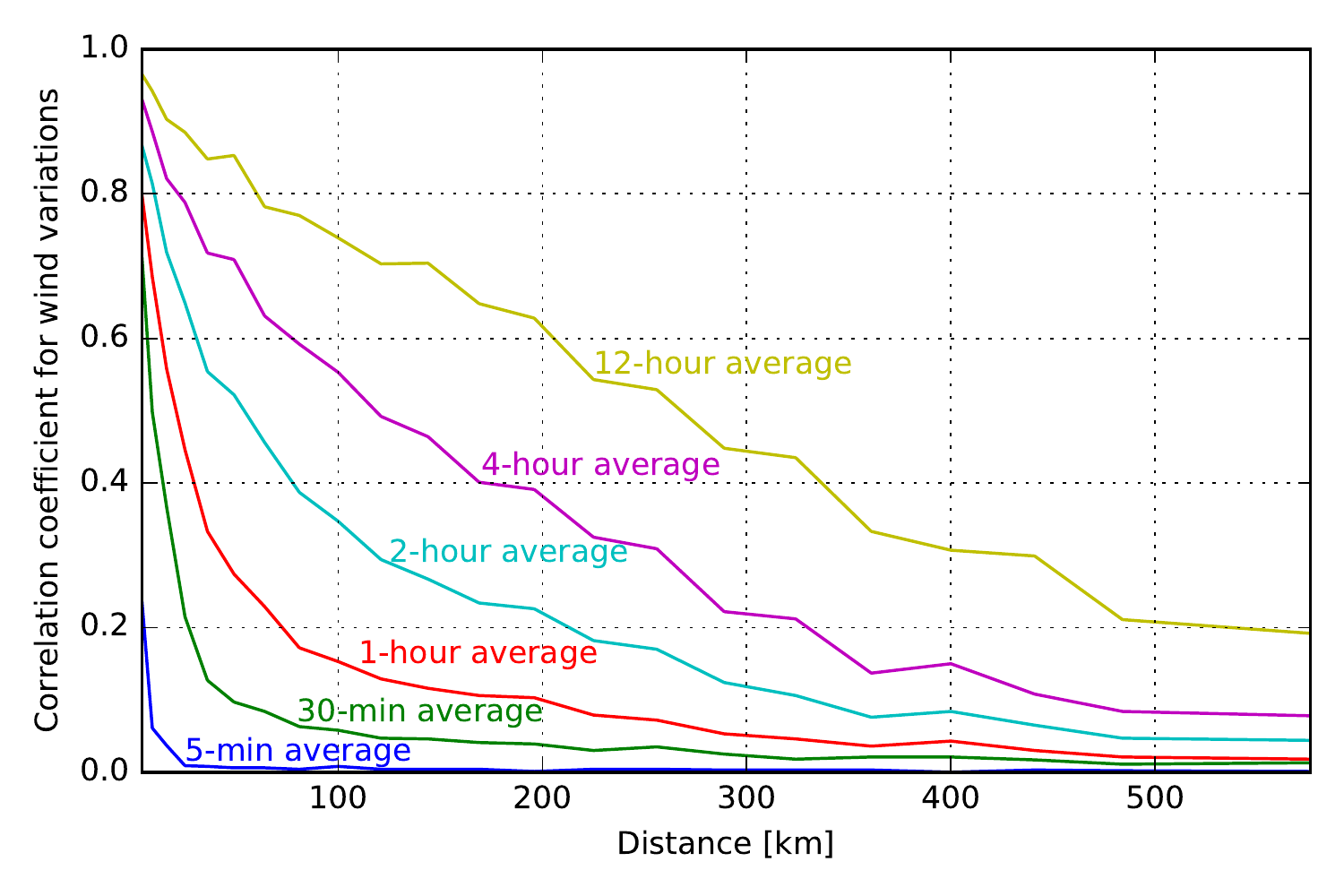}
\caption{\label{fig:correlation}Correlation of variations in wind for different time scales in Germany. (Reproduced with permission and data from \cite{ernst1999,ernst})}
\end{figure}

The reason modelling in this temporal detail is not needed is the
statistical smoothing when aggregating over a large area containing
many generators and consumers. Many of the studies are looking at the
national or sub-national level. By modelling hourly, the majority of the variation of the
demand and variable renewables like wind and solar over these areas is
captured; if there is enough flexibility to deal with the largest
hourly variations, there is enough to deal with any
intra-hour imbalance. Figure \ref{fig:correlation} shows correlations
in variations (i.e. the differences between consecutive production
values) in wind generation at different time and spatial
scales.\footnote{Note that for time series of production values
  (i.e. not the differences) the correlation does not decrease as
  rapidly as shown here and can remain high for hundreds of kilometres
  \cite{martin2015}.}  Changes within 5 minutes are uncorrelated above
25~km and therefore smooth out in the aggregation. Further analysis of
sub-hourly wind variations over large areas can be found in
\cite{norgard2004,holttinen2011}.

For solar photovoltaics (PV) the picture is similar at shorter time
scales: changes at the 5-minute level due to cloud movements are not
correlated over large areas. However, at 30 minutes to 1 hour there
are correlated changes due to the position of the sun in the sky or the passage of large-scale weather fronts. The
decrease of PV output in the evening can be captured at one-hour
resolution and there are plenty of feasible technologies available for
matching that ramping profile: flexible open-cycle gas turbines can
ramp up within 5-10 minutes, hydroelectric plants can ramp within
minutes or less, while battery storage and demand management can act within
milliseconds. For ramping down, solar and wind units can curtail their
output within seconds.

The engineering literature on sub-hourly modelling confirms these
considerations. Several studies consider the island of Ireland, which
is particularly challenging since it is an isolated synchronous area,
is only 275~km wide and has a high penetration of wind. One power
system study for Ireland with high share of wind power
\cite{DEANE2014152} varied temporal resolution between 60 minute and 5
minute intervals, and found that the 5 minute simulation results gave
system costs just 1\% higher than hourly simulation results; however,
unit commitment constraints and higher ramping and cycling rates could
be problematic for older thermal units (but not for the modern,
flexible equipment outlined above). Similarly,
\cite{6345631,ODwyer2015} see not feasibility problems at sub-hourly time
resolutions, but a higher value for flexible generation and storage,
which can act to avoid cycling stress on older thermal plants. In
\cite{DIMOULKAS2017} the difference between hourly and 15-minute
simulations in small district heating networks with high levels of
wind power penetration was considered and it was found that `the
differences in power generation are small' and `there is [no] need for
higher resolution modeling'.

To summarise, since at large spatial scales the variations in
aggregated load, wind and solar time series are statistically smoothed
out, none of the large-scale model results change significantly when
going from hourly resolution down to 5-minute simulations. Hourly
modelling will capture the biggest variations and is therefore
adequate to dimension flexibility requirements. (Reserve power and the
behaviour of the system in the seconds after faults are discussed
separately in Section \ref{sec:ancillary}.)  Sub-hourly modelling may
be necessary for smaller areas with older, inflexible thermal power
plants, but since flexible peaking plant and storage are economically
favoured in highly renewable systems, sub-hourly modelling is less
important in the long-term.

Simulations with intervals longer than one hour should be treated
carefully, depending on the research question \cite{PFENNINGER20171}.

\subsection{Their Feasibility Criterion 2b: Extreme climatic events}
\label{sec:extreme}

The authors reserve a point for studies that include rare climatic
events, such as long periods of low sun and wind, or years when
drought impacts the production of hydroelectricity.

Periods of low sun and wind in the winter longer than a few days can be
met, where available, by hydroelectricity, dispatchable biomass, demand response, imports, medium-term storage, synthetic gas from power-to-gas facilities  (the
feasibility of each of these is discussed separately below) or, in the
worst case, by fossil fuels.

From a feasibility point of view, even in the worst possible case that
enough dispatchable capacity were maintained to cover the peak load,
this does not invalidate these scenarios. The authors write "ensuring
stable supply and reliability against all plausible
outcomes\ldots{}will raise costs and complexity". Yet again, a
feasibility criterion has become a viability criterion.

So what would it cost to maintain an open-cycle gas turbine (OCGT)
fleet to cover, for example, Germany's peak demand of 80 GW?
For the OCGT we take the cost assumptions from \cite{schroeder2013}:
overnight investment cost of 400 \euro{}/kW, fixed operation and
maintenance cost of 15~\euro/kW/a, lifetime of 30 years and discount
rate of 10\%. The latter two figures given an annuity of 10.6\% of the
overnight investment cost, so the annual cost per kW is
57.4~\euro/kW/a. For a peak load of 80 GW, assuming 90\% availability
of the OCGT, the total annual cost is therefore
5.1~billion~\euro{}/a. Germany consumes more than 500~TWh/a, so this
guaranteed capacity costs less than 0.01~\euro/kWh. This is just 7.3\%
%>>> 5.1/69.4 = 0.07348703170028817
of total spending on electricity in Germany (69.4~billion~\euro{} in 2015  \cite{expertenkommission2015}).

We are not suggesting that Germany builds an OCGT fleet to cover its
peak demand. This is a worst-case rhetorical thought experiment, assuming
that no biomass, hydroelectricity, demand response, imports or medium-term storage can be activated, yet it is still low cost. Solutions that use storage that
is already in the system are likely to be even lower cost.
However, some OCGT
capacity could also be attractive for other reasons: it is a flexible
source of upward reserve power and it can be used for other ancillary services such as inertia provision, fault current, voltage regulation and black-starting the system. A clutch can even be put
on the shaft to decouple the generator from the
turbine and allow the generator to operate in synchronous compensator
mode, which means it can also provide many ancillary services without burning gas (see the discussion on ancillary services in Section \ref{sec:ancillary}).

Running the OCGT for a two-week-long low-sun-and-wind period would add
fuel costs and possibly also net \co{} emissions (which would be zero if
synthetic methane is produced with renewable energy or low if the carbon dioxide produced is captured and stored or used). Any emissions
must be accounted for in simulations, but given that extreme
climatic events are by definition rare (two weeks every decade is
0.4\% of the time; the authors even speak of once-in-100-year events),
their impact will be small.

A recent study of seven different weather years (2006 to 2012),
including extreme weather events, in Europe for a scenario with a 95\% \co{} reduction
compared to 1990 in electricity, heating and transport
\cite{IWESextreme} came to similar conclusions. The extreme events do
not affect all countries simultaneously so, for example, Germany can
cover extreme events by importing power from other countries. If for
political reasons each country is required to cover its peak load
on a national basis, the extra costs for capacity are at most 3\% of the total
system costs.

For systems that rely on hydroelectricity, the authors are right to
point out that studies should be careful to include drier years in
their simulations. Beyond the examples they cite, Brazil's hydroelectric production has been
restricted over the last couple of years due to drought, and there are periodic drier years in
Ethiopia, Kenya and Scandinavia, where in the latter inflow can drop to 30\% below the average \cite{norgard2004}.

However, in most countries, the scenarios rely on wind and solar
energy, and here the dispatchable power capacity of the hydro is
arguably just as important in balancing wind and solar as
the total yearly energy contribution, particularly if pumping can
be used to stock up the hydro reservoirs in times of wind and solar
abundance \cite{Jacobson08122015,barbosa2016}.

Note that nuclear also suffers from planned and unplanned outages, which
are exacerbated during droughts and heatwaves, when the water supplies
for river-cooled plants are either absent or too warm to provide
sufficient cooling \cite{Roehrkasten2015}. This problem is likely to
intensify given rising demand for water resources and climate change
\cite{Roehrkasten2015}.

\subsection{Their Feasibility Criterion 3: Transmission and distribution grids}
\label{sec:grid}

The authors criticise many of the studies for not providing
simulations of the transmission (i.e. high voltage long-distance grid)
and distribution (i.e. lower voltage distribution from transmission
substations to consumers) grids. Again, this is important, but not as
important as the authors assume. Feasibility is not the issue (there
are no technical restrictions on expanding the grid), but there are
socio-economic considerations. Many studies that do not model the
grid, do include blanket costs for grid expansion (e.g. from surveys
such as \cite{UEC13,HIR15,HESS2018874}).

On a cost basis, the grid is not decisive either: additional grid costs tend to
be a small fraction of total electricity system costs (examples to
follow, but typically around 10-15\% of total system costs in Europe \cite{ICNERA,RLP,Brown,energynautics,Hagspiel,Hoersch2017,HESS2018874}), and
optimal grid layouts tend to follow the cheapest generation, so
ignoring the grid is a reasonable first order approximation. Where it
can be a problem is if public acceptance problems prevent the
expansion of overhead transmission lines, in which case the power
lines have to be put underground (typically 3-8 times more expensive
than overhead lines) or electricity has to be generated more locally
(which can drive up costs and may require more storage to balance
renewables). Public acceptance problems affect cost, i.e. economic viability, not feasibility.

How much the distribution grid needs to be expanded also depends on
how much the scenario relies on decentralised, rooftop PV
generation. If all wind and utility-scale PV is connected to the
transmission grid, then there is no need to consider distribution
grids at all. Regardless of supply-side changes, distribution grids
may have to be upgraded in the future as electricity demand from
heating and electric vehicles grows (although this is not obvious:
distribution grids are often over-dimensioned for the worst possible
simultaneous peak demand, and more intelligent network infrastructure,
demand management or storage could avoid distribution grid upgrades).

Now to some examples of transmission and distribution grid costing.

A study by Imperial College, NERA and DNV GL for the European
electricity system to 2030 \cite{ICNERA} examined the consequences for
both the transmission and distribution grid of renewable energy
penetration up to 68\% (in their Scenario 1). For total annual system
costs of 232~billion~\euro{}/a in their Scenario~1, 4~billion~\euro/a
is assigned to the costs of additional transmission grid investments and 18~billion~\euro/a to the distribution grid. If there is a greater
reliance on decentralised generation (Scenario 1(a)-DG), additional
distribution grid costs could rise to 24~billion~\euro/a.

This shows a typical rule of thumb: additional grid costs are around 10-15\% of
total system costs. But this case considered only 68\% renewables.

The distribution grid study of 100\% renewables in the German federal state of Rhineland-Palatinate
(RLP) \cite{RLP} also clearly
demonstrates that the costs of generation dwarf the grid costs.
Additional grid investments vary between 10 and 15\% of the total costs of new generation, depending on how
smart the system is. Again, distribution upgrade costs dominate
transmission costs.

In its worst case the Germany Energy Agency (DENA) sees a total investment
need of 42.5 billion \euro{} in German distribution grids by 2030 for a renewables share of 82\% \cite{DENA12}.
Annualised to 4.25 billion~\euro/a, this is just 6.2\% of total spending on electricity in Germany (69.4~billion~\euro{} in 2015 \cite{expertenkommission2015}).

%Worst case is Bundesländerszenario; In dem Bundesländerszenario werden bis zum
%Jahr 2030 eine installierte EE-Leistung von 222 GW und ein EE-Anteil
%von 82 Prozent an der Bruttostromnachfrage erreicht

Another study for Germany with 100\% renewable electricity showed that
grid expansion at transmission and distribution level would cost
around 4-6~billion~\euro/a (with a big uncertainty range reaching from
1 to 12~billion~\euro/a) \cite{HESS2018874}.

Many studies look at the transmission grid only. The 2016 Ten Year
Network Development Plan (TYNDP) \cite{TYNDP2016} of the European Transmission
System Operators foresees 70-80 billion \euro{} investment needs in
Europe for 60\% renewables by 2030, which annualises to 2\% of total
electricity spending of 400~billion~\euro/a (the 0.001 to 0.002
\euro{}/kWh extra costs are compensated by a resulting reduction in
wholesale electricity prices of 0.0015 to 0.005~\euro{}/kWh
\cite{TYNDP2016}). The authors criticise the Greenpeace Energy
     [R]evolution scenario \cite{er2012,er2015} for excluding grid and
     reliability simulations, but in fact Greenpeace commissioned
     transmission expansion studies for Europe using hourly
     simulations, one for 77\% renewables by 2030 \cite{Brown} (60
     billion \euro{} investment by 2030, i.e. 1.5\% of spending) and
     one for 97\% renewables by 2050 \cite{energynautics} (149-163
     billion \euro{} investment for 97\% renewables by 2050, i.e. 4\% of
     spending). Beyond Europe, other studies with similar results look
     at the United States \cite{macdonald2017}, South and Central
     America \cite{10.1371/journal.pone.0173820}, and Asia
     \cite{en10050583,BOGDANOV2016176}.

The authors quote studies that look at optimal cross-border
transmission capacity in Europe at very high shares of renewables,
which show an expansion of 4-6 times today's capacities
\cite{Rodriguez2013,Becker2012}. It is worth pointing out that these
studies look at the international interconnectors, not the full
transmission grid, which includes the transmission lines within each
country. The interconnectors are historically weak compared to
national grids\footnote{The TYNDP \cite{TYNDP2016} will double cross-border capacities by 2030, but total circuit length will only grow by around 25\%.} and restricted by poor market design and
operation \cite{ACER2016}; if a similar methodology to \cite{Rodriguez2013,Becker2012}
is applied to a more detailed grid model with nodal pricing, the expansion is only
between 25\% and 50\% more than today's capacity
\cite{Hoersch2017}. Furthermore, cost-optimal does not necessarily
mean socially viable; there are solutions with lower grid expansion
and hence higher public acceptance, but higher storage costs to
balance renewables locally \cite{Hoersch2017}.

\subsection{Their Feasibility Criterion 4: Ancillary services}
\label{sec:ancillary}

Finally, we come to ancillary services. Ancillary services are
additional services that network operators need to stabilise and
secure the electricity system. They are mostly provided by
conventional dispatchable generators today. Ancillary services include
reserve power for balancing supply and demand in the short term,
rotating inertia to stabilise the frequency in the very short term,
synchronising torque to keep all generators rotating at the same
frequency, voltage support through reactive power provision, short
circuit current to trip protection devices during a fault, and
the ability to restart the system in the event of a total system blackout
(known as `black-starting').  The authors raise concerns that many studies do not
consider the provision of these ancillary services, particularly for
voltage and frequency control.  Again, these concerns are overblown:
ancillary services are important, but they can be provided with
established technologies (including wind and solar plants), and the cost to provide them is second order
compared to the costs of energy generation.

We consider fault current, voltage support and inertia first. These
services are mostly provided today by synchronous generators, whereas
most new wind, solar PV and storage units are coupled to the grid with
inverters, which have no inherent inertia and low fault
current, but can control voltage with both active and reactive power.

From a feasibility point of view, synchronous compensators could be
placed throughout the network and the problem is solved, although
this is not as cost effective as other solutions. Synchronous
compensators (SC), also called synchronous condensers, are
essentially synchronous generators without a prime mover to provide
active power. This means they can provide all the ancillary services
of conventional generators except those requiring active power,
i.e. they can provide fault current, inertia and voltage support just
like a synchronous generator. Active power is then provided by renewable generators and storage devices.

In fact, existing generators can be retrofitted to be SC, as happened
to the nuclear power plant in Biblis, Germany \cite{amprionbiblis}, or
to switch between generation mode and SC mode; extra mass can be added
with a clutch if more inertia is needed (SC have an inertia time
constant of 1-2~s \cite{DNVSys12,DENA16}, compared to typical
conventional generators with around 6~s). SC are a tried-and-tested
technology and have been installed recently in Germany
\cite{tennetgermany}, Denmark, Norway, Brazil, New Zealand and
California \cite{siemenssc}. They are also used in Tasmania
\cite{Tasmania2016}, where `Hydro Tasmania, TasNetworks and AEMO have
implemented many successful initiatives that help to manage and
maintain the security of a power system that has a high penetration of
asynchronous energy sources\dots Some solutions implemented in
Tasmania have been relatively low cost and without the need for
significant capital investment' \cite{Tasmania2016}.  In
Denmark, newly-installed synchronous compensators along with exchange
capacity with its neighbours allow the power system to operate without any large central power stations at all \cite{Orths2016}. In 2017 the system operated for 985 hours without central power stations, the longest continuous period of which was a week \cite{Denmark2018}. SC were also one
of the options successfully shown to improve stability during severe
faults in a study of high renewable penetration in the United States
Western Interconnection \cite{miller2015,wwsis3a}. The study concluded
`the Western Interconnection can be made to work well in the first
minute after a big disturbance with both high wind and solar and
substantial coal displacement, using good, established planning and
engineering practice and commercially available technologies'.
In a study for the
British transmission system operator National Grid \cite{urdal2015} it
was shown that 9 GVAr of SC would stabilise the British grid during
the worst fault even with 95\% instantaneous penetration of
non-synchronous generation. (Britain is tricky because it is not
synchronous with the rest of Europe and can suffer small signal
angular instability between England and Scotland.)

%XX Also: Germany, Norway, Georgia XX

%XX Mostly installed for voltage and fault current in Denmark; but for
%inertia in Tasmania and in Denmark \cite{Kroposki2017} XX

%New Zealand:
%http://www.acenz.org.nz/uploads/Events/INNOVATE%20Projects/15.%20Haywards%20Synchronous%20Condenser%20Refurbishment%20-%20webprofile-edited%20by%20HM.pdf
%https://www.energy.siemens.com/ru/pool/hq/automation/power-generation/diagnostic-suite/documents/Haywards_modernization_machinery-protection_diagnostic-suite_sppa-d3000_control-system_sppa-t3000_electrical-solutions_sppa-e3000.pdf
%Brazil: http://www.weg.net/institutional/NZ/en/news/products-and-solutions/weg-manufactures-mega-synchronous-condensers

So how cost-effective would synchronous compensators be? There is a
range of cost estimates in the literature
\cite{Kueck200627,Igbinovia2016,DNVSys12,OTHINA}, the highest being an
investment cost of 100~\euro/kVAr with fixed operating and maintenance
costs of 3.5~\euro/kVAr/a \cite{OTHINA} (it would be around a third
cheaper to retrofit existing generators \cite{DNVSys12}). For
Great Britain, the 9 GVAr of SC would cost
%9e6 * (0.106+0.035)*100) =
%GB load of 400
129 million \euro{} per year, assuming a lifetime of 30 years and a
discount rate of 10\%. That annualises to just 0.0003~\euro/kWh. (SC
also consume a small amount of active power \cite{gesc,DNVSys12}, but
given that they would run when marginal electricity costs are very low
thanks to high wind and solar feed-in, this cost would be negligible.)

Synchronous condensers are an established, mature technology, which
provide a feasible upper bound on the costs of providing
non-active-power-related ancillary services. The inverters of wind,
solar and batteries already provide reactive power for voltage control and can provide the other ancillary
services, including virtual or synthetic inertia, by programming the
functionality into the inverter software
\cite{Kroposki2017}. Inverters are much more flexible than
mechanics-bound synchronous generators and can change their output with high accuracy within
milliseconds \cite{Milligan2015}. The reason that wind and solar plants have only recently
been providing these services is that before (i.e. at lower renewable
penetration) there was no need, and no system operators required it. Now
that more ancillary services are being written into grid codes
\cite{IRENA2016}, manufacturers are providing such capabilities in their equipment.
Frequency control concepts for inverters that follow a stiff external grid frequency and adjust their active power output to compensate for any frequency deviations are already offered by manufacturers \cite{macdowell2015}.
Next generation `grid-forming' inverters will also be able to work in weak grids without a stiff frequency, albeit at the cost of increasing the inverter current rating (e.g. by 20-50\%).
A survey of different frequency-response technologies in the Irish context can be found in \cite{rocof2}.
Recent work
for National Grid \cite{strathprints58052,strathprints58053} shows
that with 25\% of inverters operating as Virtual Synchronous Machines
(VSM), the system can survive the most severe faults even when approaching 100\% non-synchronous penetration. The
literature in the control theory community on the design and stability
of grid-forming inverters in power systems is substantial and growing, and includes
both extensive simulations and tests in the field \cite{chen2011,karapanos2011,zhong2011,torres2013,bevrani2014,jouini2016,sinha2017}.

Protection systems often rely on synchronous generators to supply
fault current to trip over-current relays. Inverters are not
well-suited to providing fault current, but this can be
circumvented by replacing over-current protection with differential
protection and distance protection \cite{IRENA2015,Kroposki2017}, both
of which are established technologies.

Next, we consider balancing reserves. Balancing power can be provided
by traditional providers, battery systems, fast-acting
demand-side-management or by wind and solar generators (upward
reserves are provided by variable renewable plants by operating them below
their available power, called `delta' control). There is a wide
literature assessing requirements for balancing power with high shares
of renewables.  In a study for Germany in 2030 with 65 GW PV and 81 GW
wind (52\% renewable energy share), no need is seen for additional
primary reserve, with at most a doubling of the need for other types
of reserves \cite{DENASS}. It is a similar story in the 100\%
renewable scenario for Germany of Kombikraftwerk 2
\cite{KWK2}. (Maintaining reserves in Germany cost 315.9~million~\euro{}
in 2015 \cite{BNetzA2016}.) There is no feasibility problem here
either.

%https://www.netzentwicklungsplan.de/sites/default/files/paragraphs-files/nep_2013_1_entwurf_teil_1_kap_1_bis_8.pdf
%Page 60 energy balance
%Scenario B 2033: 567.5 TWh/a Load (with losses)
%RES: 294.2 TWh/a
%RES share: 51.8%

Another ancillary service the authors mention is black-start
capability. This is the ability to restart the electricity system in
the case of a total blackout. Most thermal power stations consume
electricity when starting up (e.g. powering pumps, fans and other
auxiliary equipment), so special provisions are needed when
black-starting the system, by making sure there are generators which
can start without an electricity supply. Typically system operators
use hydroelectric plants (which can generate as soon as the sluice
gate is opened), diesel generators or battery systems, which can then
start a gas turbine, which can then start other power plants (for
example). Maintaining conventional capacity for black-start is
inexpensive compared to system costs, as shown in Section
\ref{sec:extreme}; in a study for Germany in 2030 \cite{DENASS} with
52\% renewables, no additional measures for black-starting were deemed necessary, contrary to the interpretation in \cite{burden};
finally, decentralised renewable generators and storage could also
participate in black-starting the system in future \cite{KWK2}. The
use of battery storage systems to black-start gas turbines has
recently been demonstrated in Germany \cite{wemag-black} and in a
commercial project in California \cite{california-black}.

Nuclear, on the other hand, is a problem for black-starting, since
most designs need a power source at all times, regardless of blackout
conditions, to circulate coolant in the reactor and prevent
meltdown conditions. This will only exacerbate the need for backup generation in
a total blackout. Nuclear is sometimes not used to provide primary
reserves either, particularly in older designs, because fast changes in output present operational and safety concerns.
%E.g. AGRs in UK

\subsection{Our Feasibility Criterion 5: Fuel source that lasts more than a few decades}
\label{sec:fuel}

Here we suggest a feasibility criterion not included on the authors'
list: The technology should have a fuel source that can both supply
all the world's energy needs (not just electricity, but also
transport, heating and industrial demand) and also last more than a
couple of decades.

Traditional nuclear plants that use thermal-neutron fission of uranium do not satisfy
this feasibility criterion. In 2015 there were 7.6
million tonnes of identified uranium resources commercially recoverable at less
than 260~US\$/kgU \cite{uranium2016}.\footnote{There are further speculative and unconventional uranium resources, including in sea water, but the cost and energy required to extract them make them unviable \cite{abbott2011}.} From one tonne of
natural uranium, a light-water
reactor can generate around 40~GWh of electricity.

In 2015, world electricity consumption was around 24,000 TWh/a
\cite{iea2016}.  Assuming no rise in electricity demand and ignoring
non-electric energy consumption such as transport and
heating, uranium resources of 7.6 million tonnes will last 13 years.
%>>> 7.6*40/24. = 12.666666666666666
Reprocessing, at higher cost, might extend this by a few more years.
Including non-electric energy consumption would more than halve this time.

For renewables, exploitable energy potentials exceed yearly energy
demand by several orders of magnitude
\cite{Perez2015} and, by definition, are not depleted over time. Even
taking account of limitations of geography and material resources, the
potentials for the expansion of wind, solar and storage exceed
demand projections by several factors \cite{Jacobson2011a}.

As for `following all paths' and pursuing a mix of renewables and
nuclear, they do not mix well: because of their high capital costs,
nuclear power plants are most economically viable when operated at
full power the whole time, whereas the variability of renewables
requires a flexible balancing power fleet \cite{VERBRUGGEN20084036}. Network expansion can help the
penetration of both renewables and inflexible plant \cite{HAMACHER2013657}, but this would create further
pressure for grid expansion, which is already pushing against social
limits in some regions.

This feasibility criterion is not met by standard nuclear reactors, but
could be met in theory by breeder reactors and fusion power. This
brings us to our next feasibility criterion.

\subsection{Our Feasibility Criterion 6: Should not rely on unproven technologies}
\label{sec:unproven}

Here is another feasibility criterion that is not included on the
authors' list: Scenarios should not rely on unproven technologies.  We
are not suggesting that we should discontinue research into new
technologies, rather that when planning for the future, we should be
cautious and assume that not every new technology will reach technical
and commercial maturity.

The technologies required for renewable scenarios are not just
tried-and-tested, but also proven at a large scale. Wind, solar, hydro
and biomass all have capacity in the hundreds of GWs worldwide \cite{REN17}. The necessary
expansion of the grid and ancillary services can deploy existing
technology (see Sections \ref{sec:grid} and \ref{sec:ancillary}). Heat pumps are used widely \cite{en10040578}. Battery storage, contrary to the authors' paper, is a
proven technology already implemented in billions of devices
worldwide (including a utility-scale 100~MW plant in South Australia \cite{Tesla100MW2017} and 700~MW of utility-scale batteries in the United States at the end of 2017 \cite{EIA2018}). Compressed air energy storage, thermal storage, gas storage, hydrogen electrolysis, methanation and fuel cells are all decades-old
technologies that are well understood. (See Section \ref{sec:storage} for more on the feasibility of storage technologies.)

On the nuclear side, for the coming decades when uranium for
thermal-neutron reactors would run out, we have breeder reactors,
which can breed more fissile material from natural uranium or thorium,
or fusion power.

Breeder reactors are technically immature (with a technology readiness
level between 3 and 5 depending on the design \cite{INL2015}), more
costly than light-water reactors, unreliable, potentially unsafe and
they pose serious proliferation risks \cite{cochran2010}.  Most
fast-neutron breeder reactors rely on sodium as a coolant, and
since sodium burns in air and water, it makes refueling and repair
difficult. This has led to serious accidents in fast breeder reactors,
such as the major sodium fire at the Monju plant in 1995. Some experts
consider fast breeders to have already failed as a technology option
\cite{cochran2010,slowdeath}. The burden
of proof is on the nuclear industry to demonstrate that breeder reactors
are a safe and commercially competitive technology.

Fusion power is even further from demonstrating technical feasibility.
No fusion plant exists today that can generate more energy than it
requires to initiate and sustain fusion. Containment materials that can withstand the neutron bombardment without generating long-lived nuclear waste are still under development. Even advocates of fusion do
not expect the first commercial plant to go online before 2050
\cite{EFDA}. Even if it proves to be feasible and cost-effective
(which is not clear at this point), ramping up to a high worldwide
penetration will take decades more. That is too late to tackle global
warming \cite{IPCC2014-III-SPM}.

\section{Other Issues}\label{sec:other}

In this section we address other issues raised by the authors of
\cite{burden} during their discussion of their feasibility criteria.

\subsection{Feasibility of storage technologies}
\label{sec:storage}

The authors write "widespread storage of energy using a range of
technologies (most of which - beyond pumped hydro - are unproven at
large scales, either technologically and/or economically)".

Regarding battery storage, it is clear that there is the potential to
exploit established lithium ion technology at scale and at low cost
\cite{Nykvist2015,schmidt2017,kittner2017}. The technology is already
widely established in electronic devices and increasingly in battery
electric vehicles, which will in future provide a regular and cheap
source of second-life stationary batteries.  A utility-scale 100~MW plant was
installed in the South Australian grid in 2017 \cite{Tesla100MW2017} and there
was already 700~MW of utility-scale batteries in the United States at the end of
2017 \cite{EIA2018}.  Further assessments of the potential for lithium
ion batteries can be found in \cite{Jacobson2011a}. Costs are falling
so fast that hybrid PV-battery systems are already or soon will be
competitive with conventional systems in areas with good solar
resources \cite{AFANASYEVA2016157,GTM2018}.

Many other electricity storage devices have been not just demonstrated
but already commercialised \cite{luo2015}, including large-scale
compressed air energy storage. Technologies that convert electricity
to gas, by electrolysing hydrogen with the possibility of later
methanation, are already being demonstrated at megawatt scale
\cite{gotz2016,RONSCH2016276}. Hydrogen could either be fed into the
gas network to a certain fraction, used in fuel cell vehicles,
converted to other synthetic fuels, or converted back into electricity
for the grid. Fuel cells are already manufactured at gigawatt scale,
with 480~MW installed in 2016 \cite{FuelCell16}. By using the process
heat from methanation to cover the heat consumption of electrolysis,
total efficiency for power-to-methane of 76\% has recently been
demonstrated in a freight-container-sized pilot project, with 80\% efficiency in sight \cite{HELMETH}.

Moreover, in a holistic, cross-sectoral energy systems approach that
goes beyond electricity to integrate all thermal, transport and
industrial demand, it is possible to identify renewable energy systems
in which all storage is based on low-cost well-proven technologies,
such as thermal, gas and liquid storage, all of which are cheaper than
electricity storage \cite{IJSEPM1574}. These sectors also provide
significant deferrable demand, which further helps to integrate
variable renewable energy \cite{4957254,MATHIESEN2015139,Brown2018}.
Storage capacity for natural gas in the European Union is 1075~TWh as
of mid 2017 \cite{gie}.

\subsection{Feasibility of biomass}
\label{sec-5-3}

The authors criticise a few studies for their over-reliance on
biomass, such as one for Denmark \cite{LUND2009524} and one for
Ireland \cite{CONNOLLY2011502}. There are legitimate concerns about
the availability of fuel crops, environmental damage, biodiversity loss and competition
with food crops \cite{GCBB:GCBB12205}. More recent studies, including some by the same
researchers, conduct detailed potential assessments for biomass and/or
restrict biomass usage to agricultural residues and waste
\cite{LINDFELDT20101836,MATHIESEN2012160,LUND2015389,Connolly20161634}.  Other studies are even
more conservative (or concerned about air pollution from combustion products \cite{millstein2017}) and exclude biomass altogether
\cite{Schlachtberger2017,Jacobson2011a,Jacobson08122015,Jacobson2017,Brown2018}, while still reaching feasible and
cost-effective energy systems.

\subsection{Feasibility of carbon capture}

Capturing carbon dioxide from industrial processes, power plants or
directly from the air could also contribute to mitigating net
greenhouse gas emissions. The captured carbon dioxide can then be used
in industry (e.g. in greenhouses or in the production of synthetic
fuels) or sequestered (e.g. underground). While some of the individual
components have been demonstrated at commercial scale, hurdles \cite{Pires2011,Boots2014,LEUNG2014426} include
cost, technical feasibility of long-term sequestration without leakage,
viability for some concepts (such as direct air capture (DAC), the lowest cost
version of which is rated at Technology Readiness Level (TRL) 3-5 \cite{McLaren2012}), other air
pollutants from combustion and imperfect capture when capturing from power plants, lower energy efficiency, regulatory issues, public
acceptance of sequestration facilities \cite{LORANGESEIGO2014848} and systems integration.

Studies at high time resolution that have combined renewables and power plants with carbon capture and sequestration (CCS)
suggest that CCS is not cost effective because of high
capital costs and low utilisation \cite{LUND2012469}. However, DAC may
be promising for the production of synthetic fuels \cite{MATHIESEN2015139,FASIHI2016243,su9020306} and is attractive because of its locational flexibility and minimal water consumption \cite{Keith1654,Williamson2016}. Negative emissions technologies (NET), which include DAC, bioenergy with CCS, enhanced weathering, ocean fertilisation, afforestation  and reforestation, may also be necessary to meet the goals of the Paris climate accord \cite{Fuss2014,Schleussner2016,vanVuuren2017,Rockstrom1269}. Relying on NET presents risks given their technical immaturity, so further research and development of these technologies is required \cite{Fuss2014,Smith2015,Anderson2016,Vaughan2016}.

\subsection{Viability of renewable energy systems}\label{sec:re_viable}

In the sections above we have shown that energy systems with very high
shares of renewable energy are both feasible and economically viable
with respect to primary energy demand projections, matching short-term
variability, extreme events, transmission and distribution grids,
ancillary services, resource availability and technological maturity.
We now turn to more general points of social and economic viability.

With regard to social viability, there are high levels of public
support for renewable energy. In a survey of European Union citizens
for the European Commission in 2017, 89\% thought it was important for
their national government to set targets to increase renewable energy
use by 2030 \cite{EUsurvey2017}. A 2017 survey of the citizens of 13
countries from across the globe found that 82\% believe it is
important to create a world fully powered by renewable
energy \cite{Orsted2017}.  A 2016 compilation of surveys from leading
industrialised countries showed support for renewables in most cases
to be well over 80\% \cite{AEE2016}.  Concerns have been raised
primarily regarding the public acceptance of onshore wind turbines and
overhead transmission lines. Repeated studies have shown that public
acceptance of onshore wind can be increased if local communities are
engaged early in the planning process, if their concerns are addressed
and if they are given a stake in the project outcome
\cite{MARUYAMA20072761,JOBERT20072751,ENEVOLDSEN2016178}.  Where
onshore wind is not socially viable, there are system solutions with
higher shares of offshore wind and solar energy, but they may cost
fractionally more \cite{Schlachtberger2018}. The picture is similar
with overhead transmission lines: more participatory governance early
in the planning stages and local involvement if the project is built
can increase public acceptance \cite{COHEN20144,GALVIN2018114}. Again,
if overhead transmission is not viable, there are system solutions
with more storage and underground cables, but they are more expensive
\cite{Hoersch2017}. The use of open data and open model software can help to
improve transparency \cite{PFENNINGER2017211,PFENNINGER201863,Child2018321}.

Next we turn to the economic viability of bulk energy generation from
renewable sources. On the basis of levelised cost, onshore wind,
offshore wind, solar PV, hydroelectricity and biomass are already
either in the range of current fossil fuel generation or lower cost
\cite{IRENA2018}. Levelised cost is only a coarse measure
\cite{Joskow2011}, since it does not take account of variability,
which is why integration studies typically consider total system costs
in models with high spatial and temporal resolution. Despite often using
conservative cost assumptions, integration studies repeatedly show
that renewables-based systems are possible with costs that are
comparable or lower than conventional fossil-fuel-based systems
\cite{Cochran2014,Jacobson2011a,Jacobson2011b,wwf,er2012,Jacobson08122015,ELLISTON2012606,Elliston,LUND2009524,CONNOLLY2011502,PWC2010,budischak2013,GROSSMANN2013831,PLEMANN201422,BOGDANOV2016176,HUBER2015180,Czisch2005,STEINKE2013826,RASMUSSEN2012642,Hagspiel,Connolly20161634,BUSSAR20161,DOMINKOVIC20161517,GROSSMANN2014983,HUBER2015235,BECKER2014443,CHILD2016517,MATHIESEN2015139,Palzer20141019,IJSEPM497,FERNANDES201451,MOELLER201439,PIP:PIP2885,PIP:PIP2950,Jacobson2017,JACOBSON2018,su9020233,en10050583,PLEMANN201719,Schlachtberger2017,Hoersch2017,GILS2017173,ERIKSEN2017913,CEBULLA2017211,Brown2018,PLEMANN2017,10.1371/journal.pone.0173820,en10081171,GULAGI2017,10.1371/journal.pone.0180611,BLAKERS2017471,LU2017663,barbosa2016,Aghahosseini2017,SADIQA2018518,Caldera2018,KILICKAPLAN2017218,CHILD2017410,GILS2017342,CHILD201749},
even before aspects such as climate impact and health outcomes are
considered.

For example, focusing on results of our own research, a global switch
to 100\% renewable electricity by 2050 would see a drop in average
system cost from 70~\euro/MWh in 2015 to 52~\euro/MWh in 2050 \cite{PIP:PIP2950}. This study modelled the electricity system at hourly resolution for an entire year for 145 regions of the world. Considering all energy sectors in Europe, costs in a 100\% renewable energy scenario would be only 10\% higher than a business-as-usual scenario for 2050 \cite{Connolly20161634}.

The low cost of renewables is borne out in recent auctions, where, for example, extremely low prices have been seen for systems that include storage in the United States due to come online in 2023 (a median PV-plus-battery price of 36~US\$/MWh and a median wind-plus-storage price of 21~US\$/MWh \cite{GTM2018}).

\subsection{Viability of nuclear power}
\label{sec-5-1}

Following the authors, we have focussed above on the technical
feasibility of nuclear. For discussions of the socio-economic viability
of nuclear power, i.e. the cost, safety, decomissioning, waste disposal, public acceptance, terrorism and
nuclear-weapons-proliferation issues resulting from current designs, see
for example \cite{B809990C,Jacobson2011a,abbott2011,VERBRUGGEN201416}.

\subsection{Other studies of 100\% renewable systems}\label{sec:other_studies}

At the time the authors submitted their article there were many other
studies of 100\% or near-100\% renewable systems that the authors did not
review. Most studies were simulated with an hourly resolution and many
modelled the transmission grid, with examples covering the globe
\cite{GROSSMANN2013831,PLEMANN201422}, North-East Asia
\cite{BOGDANOV2016176}, the Association of South-East Asian Nations
(ASEAN) \cite{HUBER2015180}, Europe and its neighbours \cite{Czisch2005}, Europe
\cite{STEINKE2013826,RASMUSSEN2012642,Hagspiel,Connolly20161634,BUSSAR20161}, South-East Europe \cite{DOMINKOVIC20161517},
the Americas \cite{GROSSMANN2014983}, China \cite{HUBER2015235}, the
United States \cite{BECKER2014443}, Finland \cite{CHILD2016517},
Denmark \cite{MATHIESEN2015139}, Germany \cite{Palzer20141019},
Ireland \cite{IJSEPM497}, Portugal \cite{FERNANDES201451} and Berlin-Brandenburg in Germany
\cite{MOELLER201439}.

Since then other 100\% studies have considered the globe
\cite{PIP:PIP2885,PIP:PIP2950,Jacobson2017,JACOBSON2018}, Asia \cite{su9020233}, Southeast Asia and the
Pacific Rim \cite{en10050583}, Europe
\cite{PLEMANN201719,Schlachtberger2017,Hoersch2017,GILS2017173,ERIKSEN2017913,CEBULLA2017211,Brown2018},
South-East Europe \cite{PLEMANN2017}, South and Central America
\cite{10.1371/journal.pone.0173820}, North America \cite{en10081171},
India and its neighbours \cite{GULAGI2017,10.1371/journal.pone.0180611}, Australia
\cite{BLAKERS2017471,LU2017663}, Brazil
\cite{barbosa2016}, Iran \cite{Aghahosseini2017}, Pakistan \cite{SADIQA2018518}, Saudi Arabia \cite{Caldera2018}, Turkey \cite{KILICKAPLAN2017218}, Ukraine \cite{CHILD2017410} the Canary Islands \cite{GILS2017342}
and the Åland Islands \cite{CHILD201749}.

\subsection{Places already at or close to 100\% renewables}
\label{sec-5-2}

The authors state that the only developed nation with 100\% renewable
electricity is Iceland. This statement ignores countries which come
close to 100\% and smaller island systems which are already at 100\%
(on islands the integration of renewables is harder, because they
cannot rely on their neighbours for energy trading or frequency
stability), which the authors of \cite{burden} chose to exclude from
their study.

Countries which are close to 100\% renewable electricity include Paraguay (99\%), Norway
(97\%), Uruguay (95\%), Costa Rica (93\%), Brazil (76\%) and Canada
(62\%) \cite{Kroposki2017}. Regions within countries which are at or
above 100\% include Mecklenburg-Vorpommern in Germany,
Schleswig-Hostein in Germany, South Island in New Zealand, Orkney in
Scotland and Samsø along with many other parts of Denmark.

This list mostly contains examples where there is sufficient
synchronous generation to stabilise the grid, either from
hydroelectricity, geothermal or biomass, or an alternating current
connection to a neighbour. There are also purely inverter-based
systems on islands in the South
Pacific (Tokelau \cite{sma2012} and an island in American Samoa) which have solar
plus battery systems. We could also include here any
residential solar plus battery off-grid systems.

Another relevant example is the German offshore collector grids in the
North Sea, which only have inverter-based generators and
consumption. Inverter-interfaced wind turbines are connected with an
alternating current grid to an AC-DC converter station, which feeds
the power onto land through a High Voltage Direct Current cable. There
is no synchronous machine in the offshore grid to stabilise it, but
they work just fine (after teething problems with unwanted
harmonics between the inverters).

Off-planet, there is also the International Space Station and other
space probes which rely on solar energy.

\subsection{South Australian blackout in September 2016}
\label{sec-5-6}

The authors implicitly blame wind generation for the South Australian
blackout in September 2016, where some wind turbines disconnected
after multiple faults when tornadoes simultaneously damaged two
transmission lines (an extreme event). According to the final
report by the Australian Energy Market Operator (AEMO) on the incident \cite{AEMO2017}
"Wind turbines successfully rode through grid disturbances. It was the
action of a control setting responding to multiple disturbances that
led to the Black System. Changes made to turbine control settings
shortly after the event [have] removed the risk of recurrence given the
same number of disturbances." AEMO still highlights the need for
additional frequency control services, which can be provided  at low cost, as
outlined in Section \ref{sec:ancillary}.

\section{Conclusions}\label{sec:conclusions}

In `Burden of proof: A comprehensive review of the feasibility of
100\% renewable-electricity systems' \cite{burden} the authors called
into question the feasibility of highly renewable scenarios.  To
assess a selection of relevant studies, they chose feasibility criteria
that are important, but not critical for either the feasibility or
viability of the studies. We have shown here that all the issues can
be addressed at low economic cost. Worst-case, conservative technology
choices (such as dispatchable capacity for the peak load, grid
expansion and synchronous compensators for ancillary services) are not
only technically feasible, but also have costs which are a magnitude
smaller than the total system costs. More cost-effective solutions
that use variable renewable generators intelligently are also
available. The viability of these solutions justifies the focus of
many studies on reducing the main costs of bulk energy generation.

As a result, we conclude that the 100\% renewable energy scenarios proposed in
the literature are not just feasible, but also viable. As we demonstrated in
Section \ref{sec:re_viable}, 100\% renewable systems that meet the energy needs of all citizens at all times are
cost-competitive with fossil-fuel-based systems, even before
externalities such as global warming, water usage and environmental
pollution are taken into account.

The authors claim that a 100\% renewable world will require a
`re-invention' of the power system; we have shown here that this claim
is exaggerated: only a directed evolution of the current system is
required to guarantee affordability, reliability and sustainability.

\section*{Acknowledgements}

We thank Thomas Ackermann, Florian Dörfler, Ben Elliston, Veit
Hagenmeyer, Bri-Mathias Hodge, Jonas Hörsch, Nick Miller, Robbie
Morrison, Antje Orths, Mirko Schäfer, David Schlachtberger, Charlie
Smith, Bo Tranberg and Helge Urdal for helpful discussions,
suggestions and comments.  For their permission to reproduce the
figures we thank the German Federal Ministry for Economic Affairs and
Energy (Figure \ref{fig:weissbuch}) and Bernhard Ernst
(Figure \ref{fig:correlation}).  T.B. acknowledges funding from the
Helmholtz Association under grant no.~VH-NG-1352 and from the German
Federal Ministry of Education and Research (BMBF) under grant
no.~03SF0472C. T.B., H.L. and B.V.M. acknowledge funding from the
RE-INVEST project under grant number 6154-00022B from the Innovation
Fund Denmark.  The responsibility for the contents lies with the
authors.

\bibliographystyle{elsarticle-num}

%Tip from https://tex.stackexchange.com/questions/82293/how-to-generate-multiple-numeric-style-citation-call-outs
\biboptions{sort&compress}
\bibliography{burden-response}

\begin{thebibliography}{100}
\expandafter\ifx\csname url\endcsname\relax
  \def\url#1{\texttt{#1}}\fi
\expandafter\ifx\csname urlprefix\endcsname\relax\def\urlprefix{URL }\fi
\expandafter\ifx\csname href\endcsname\relax
  \def\href#1#2{#2} \def\path#1{#1}\fi

\bibitem{IPCC2014-synthesis}
{Core Writing Team}, R.~Pachauri, L.~Meyer (Eds.),
  \href{https://www.ipcc.ch/report/ar5/syr/}{{Climate Change 2014: Synthesis
  Report. Contribution of Working Groups I, II and III to the Fifth Assessment
  Report of the Intergovernmental Panel on Climate Change}}, Cambridge
  University Press, Cambridge, United Kingdom and New York, NY, USA, 2014.
\newline\urlprefix\url{https://www.ipcc.ch/report/ar5/syr/}

\bibitem{Cochran2014}
J.~Cochran, T.~Mai, M.~Bazilian,
  \href{https://doi.org/10.1016/j.rser.2013.08.089}{Meta-analysis of high
  penetration renewable energy scenarios}, Renewable and Sustainable Energy
  Reviews 29 (2014) 246 -- 253.
\newblock \href {http://dx.doi.org/10.1016/j.rser.2013.08.089}
  {\path{doi:10.1016/j.rser.2013.08.089}}.
\newline\urlprefix\url{https://doi.org/10.1016/j.rser.2013.08.089}

\bibitem{Jacobson2011a}
M.~Jacobson, M.~Delucchi, Providing all global energy with wind, water, and
  solar power, {Part I}: Technologies, energy resources, quantities and areas
  of infrastructure, and materials, Energy Policy 39~(3) (2011) 1154--1169.
\newblock \href {http://dx.doi.org/10.1016/j.enpol.2010.11.040}
  {\path{doi:10.1016/j.enpol.2010.11.040}}.

\bibitem{Jacobson2011b}
M.~Delucchi, M.~Jacobson, Providing all global energy with wind, water, and
  solar power, {Part II}: Reliability, system and transmission costs, and
  policies, Energy Policy 39~(3) (2011) 1170--1190.
\newblock \href {http://dx.doi.org/10.1016/j.enpol.2010.11.045}
  {\path{doi:10.1016/j.enpol.2010.11.045}}.

\bibitem{wwf}
The energy report: 100\% renewable energy by 2050, Tech. rep., World Wildlife
  Fund (WWF),
  \url{http://www.ecofys.com/files/files/ecofys-wwf-2011-the-energy-report.pdf}
  (2011).

\bibitem{er2012}
S.~Teske, J.~Muth, S.~Sawyer, T.~Pregger, S.~Simon, T.~Naegler, M.~O'Sullivan,
  energy [r]evolution: a sustainable world energy outlook, 4th edition, Tech.
  rep., Greenpeace International, European Renewable Energy Council (EREC),
  Global Wind Energy Council (GWEC) (2012).

\bibitem{Jacobson08122015}
M.~Z. Jacobson, M.~A. Delucchi, M.~A. Cameron, B.~A. Frew, Low-cost solution to
  the grid reliability problem with 100\% penetration of intermittent wind,
  water, and solar for all purposes, Proceedings of the National Academy of
  Sciences 112~(49) (2015) 15060--15065.
\newblock \href {http://dx.doi.org/10.1073/pnas.1510028112}
  {\path{doi:10.1073/pnas.1510028112}}.

\bibitem{ELLISTON2012606}
B.~Elliston, M.~Diesendorf, I.~MacGill,
  \href{https://doi.org/10.1016/j.enpol.2012.03.011}{{Simulations of scenarios
  with 100\% renewable electricity in the Australian National Electricity
  Market}}, Energy Policy 45 (2012) 606 -- 613.
\newblock \href {http://dx.doi.org/10.1016/j.enpol.2012.03.011}
  {\path{doi:10.1016/j.enpol.2012.03.011}}.
\newline\urlprefix\url{https://doi.org/10.1016/j.enpol.2012.03.011}

\bibitem{Elliston}
B.~Elliston, I.~MacGill, M.~Diesendorf, Least cost 100\% renewable electricity
  scenarios in the {Australian National Electricity Market}, Energy Policy
  59~(0) (2013) 270 -- 282.
\newblock \href {http://dx.doi.org/10.1016/j.enpol.2013.03.038}
  {\path{doi:10.1016/j.enpol.2013.03.038}}.

\bibitem{LUND2009524}
H.~Lund, B.~Mathiesen, {Energy system analysis of 100\% renewable energy
  systems - The case of Denmark in years 2030 and 2050}, Energy 34~(5) (2009)
  524 -- 531, 4th Dubrovnik Conference.
\newblock \href {http://dx.doi.org/10.1016/j.energy.2008.04.003}
  {\path{doi:10.1016/j.energy.2008.04.003}}.

\bibitem{CONNOLLY2011502}
D.~Connolly, H.~Lund, B.~Mathiesen, M.~Leahy, {The first step towards a 100\%
  renewable energy-system for Ireland}, Applied Energy 88~(2) (2011) 502 --
  507, the 5th Dubrovnik Conference on Sustainable Development of Energy, Water
  and Environment Systems, held in Dubrovnik September/October 2009.
\newblock \href {http://dx.doi.org/10.1016/j.apenergy.2010.03.006}
  {\path{doi:10.1016/j.apenergy.2010.03.006}}.

\bibitem{PWC2010}
G.~Schellekens, A.~Battaglini, J.~Lilliestam, J.~McDonnell, A.~Patt, {100\%
  renewable electricity: A roadmap to 2050 for Europe and North Africa}, Tech.
  rep., PriceWaterhouseCoopers (2010).

\bibitem{budischak2013}
C.~Budischak, D.~Sewell, H.~Thomson, L.~Mach, D.~E. Veron, W.~Kempton,
  Cost-minimized combinations of wind power, solar power and electrochemical
  storage, powering the grid up to 99.9\% of the time, Journal of Power Sources
  225 (2013) 60 -- 74.
\newblock \href {http://dx.doi.org/10.1016/j.jpowsour.2012.09.054}
  {\path{doi:10.1016/j.jpowsour.2012.09.054}}.

\bibitem{GROSSMANN2013831}
W.~D. Grossmann, I.~Grossmann, K.~W. Steininger,
  \href{https://doi.org/10.1016/j.rser.2012.08.018}{{Distributed solar
  electricity generation across large geographic areas, Part I: A method to
  optimize site selection, generation and storage}}, Renewable and Sustainable
  Energy Reviews 25 (2013) 831 -- 843.
\newblock \href {http://dx.doi.org/10.1016/j.rser.2012.08.018}
  {\path{doi:10.1016/j.rser.2012.08.018}}.
\newline\urlprefix\url{https://doi.org/10.1016/j.rser.2012.08.018}

\bibitem{PLEMANN201422}
G.~Pleßmann, M.~Erdmann, M.~Hlusiak, C.~Breyer,
  \href{https://doi.org/10.1016/j.egypro.2014.01.154}{{Global Energy Storage
  Demand for a 100\% Renewable Electricity Supply}}, Energy Procedia 46 (2014)
  22 -- 31, 8th International Renewable Energy Storage Conference and
  Exhibition (IRES 2013).
\newblock \href {http://dx.doi.org/10.1016/j.egypro.2014.01.154}
  {\path{doi:10.1016/j.egypro.2014.01.154}}.
\newline\urlprefix\url{https://doi.org/10.1016/j.egypro.2014.01.154}

\bibitem{BOGDANOV2016176}
D.~Bogdanov, C.~Breyer,
  \href{https://doi.org/10.1016/j.enconman.2016.01.019}{{North-East Asian Super
  Grid for 100\% renewable energy supply: Optimal mix of energy technologies
  for electricity, gas and heat supply options}}, Energy Conversion and
  Management 112 (2016) 176 -- 190.
\newblock \href {http://dx.doi.org/10.1016/j.enconman.2016.01.019}
  {\path{doi:10.1016/j.enconman.2016.01.019}}.
\newline\urlprefix\url{https://doi.org/10.1016/j.enconman.2016.01.019}

\bibitem{HUBER2015180}
M.~Huber, A.~Roger, T.~Hamacher,
  \href{https://doi.org/10.1016/j.energy.2015.04.065}{Optimizing long-term
  investments for a sustainable development of the asean power system}, Energy
  88 (2015) 180 -- 193.
\newblock \href {http://dx.doi.org/10.1016/j.energy.2015.04.065}
  {\path{doi:10.1016/j.energy.2015.04.065}}.
\newline\urlprefix\url{https://doi.org/10.1016/j.energy.2015.04.065}

\bibitem{Czisch2005}
G.~Czisch,
  \href{https://kobra.bibliothek.uni-kassel.de/bitstream/urn:nbn:de:hebis:34-200604119596/1/DissVersion0502.pdf}{{Szenarien
  zur zuk\"unftigen {S}tromversorgung: Kostenoptimierte Variationen zur
  Versorgung Europas und seiner Nachbarn mit Strom aus erneuerbaren Energien}},
  Ph.D. thesis, Universit\"at Kassel (2005).
\newline\urlprefix\url{https://kobra.bibliothek.uni-kassel.de/bitstream/urn:nbn:de:hebis:34-200604119596/1/DissVersion0502.pdf}

\bibitem{STEINKE2013826}
F.~Steinke, P.~Wolfrum, C.~Hoffmann,
  \href{https://doi.org/10.1016/j.renene.2012.07.044}{{Grid vs. storage in a
  100\% renewable Europe}}, Renewable Energy 50 (2013) 826 -- 832.
\newblock \href {http://dx.doi.org/10.1016/j.renene.2012.07.044}
  {\path{doi:10.1016/j.renene.2012.07.044}}.
\newline\urlprefix\url{https://doi.org/10.1016/j.renene.2012.07.044}

\bibitem{RASMUSSEN2012642}
M.~G. Rasmussen, G.~B. Andresen, M.~Greiner,
  \href{10.1016/j.enpol.2012.09.009}{{Storage and balancing synergies in a
  fully or highly renewable pan-European power system}}, Energy Policy 51
  (2012) 642 -- 651, renewable Energy in China.
\newblock \href {http://dx.doi.org/https://doi.org/10.1016/j.enpol.2012.09.009}
  {\path{doi:https://doi.org/10.1016/j.enpol.2012.09.009}}.
\newline\urlprefix\url{10.1016/j.enpol.2012.09.009}

\bibitem{Hagspiel}
S.~Hagspiel, C.~J\"agemann, D.~Lindenburger, T.~Brown, S.~Cherevatskiy,
  E.~Tr\"oster,
  \href{https://doi.org/10.1016/j.energy.2014.01.025}{Cost-optimal power system
  extension under flow-based market coupling}, Energy 66 (2014) 654--666.
\newblock \href {http://dx.doi.org/10.1016/j.energy.2014.01.025}
  {\path{doi:10.1016/j.energy.2014.01.025}}.
\newline\urlprefix\url{https://doi.org/10.1016/j.energy.2014.01.025}

\bibitem{Connolly20161634}
D.~Connolly, H.~Lund, B.~Mathiesen,
  \href{https://doi.org/10.1016/j.rser.2016.02.025}{{Smart Energy Europe: The
  technical and economic impact of one potential 100\% renewable energy
  scenario for the European Union}}, Renewable and Sustainable Energy Reviews
  60 (2016) 1634 -- 1653.
\newblock \href {http://dx.doi.org/10.1016/j.rser.2016.02.025}
  {\path{doi:10.1016/j.rser.2016.02.025}}.
\newline\urlprefix\url{https://doi.org/10.1016/j.rser.2016.02.025}

\bibitem{BUSSAR20161}
C.~Bussar, P.~Stöcker, Z.~Cai, L.~M. Jr., D.~Magnor, P.~Wiernes, N.~van
  Bracht, A.~Moser, D.~U. Sauer,
  \href{https://doi.org/10.1016/j.est.2016.02.004}{{Large-scale integration of
  renewable energies and impact on storage demand in a European renewable power
  system of 2050—Sensitivity study}}, Journal of Energy Storage 6 (2016) 1 --
  10.
\newblock \href {http://dx.doi.org/10.1016/j.est.2016.02.004}
  {\path{doi:10.1016/j.est.2016.02.004}}.
\newline\urlprefix\url{https://doi.org/10.1016/j.est.2016.02.004}

\bibitem{DOMINKOVIC20161517}
D.~Dominković, I.~Bačeković, B.~Ćosić, G.~Krajačić, T.~Pukšec,
  N.~Duić, N.~Markovska,
  \href{https://doi.org/10.1016/j.apenergy.2016.03.046}{Zero carbon energy
  system of south east europe in 2050}, Applied Energy 184 (2016) 1517 -- 1528.
\newblock \href {http://dx.doi.org/10.1016/j.apenergy.2016.03.046}
  {\path{doi:10.1016/j.apenergy.2016.03.046}}.
\newline\urlprefix\url{https://doi.org/10.1016/j.apenergy.2016.03.046}

\bibitem{GROSSMANN2014983}
W.~D. Grossmann, I.~Grossmann, K.~W. Steininger,
  \href{https://doi.org/10.1016/j.rser.2014.01.003}{{Solar electricity
  generation across large geographic areas, Part II: A Pan-American energy
  system based on solar}}, Renewable and Sustainable Energy Reviews 32 (2014)
  983 -- 993.
\newblock \href {http://dx.doi.org/10.1016/j.rser.2014.01.003}
  {\path{doi:10.1016/j.rser.2014.01.003}}.
\newline\urlprefix\url{https://doi.org/10.1016/j.rser.2014.01.003}

\bibitem{HUBER2015235}
M.~Huber, C.~Weissbart, \href{https://doi.org/10.1016/j.energy.2015.05.146}{{On
  the optimal mix of wind and solar generation in the future Chinese power
  system}}, Energy 90 (2015) 235 -- 243.
\newblock \href {http://dx.doi.org/10.1016/j.energy.2015.05.146}
  {\path{doi:10.1016/j.energy.2015.05.146}}.
\newline\urlprefix\url{https://doi.org/10.1016/j.energy.2015.05.146}

\bibitem{BECKER2014443}
S.~Becker, B.~A. Frew, G.~B. Andresen, T.~Zeyer, S.~Schramm, M.~Greiner, M.~Z.
  Jacobson, \href{https://doi.org/10.1016/j.energy.2014.05.067}{{Features of a
  fully renewable US electricity system: Optimized mixes of wind and solar PV
  and transmission grid extensions}}, Energy 72 (2014) 443 -- 458.
\newblock \href {http://dx.doi.org/10.1016/j.energy.2014.05.067}
  {\path{doi:10.1016/j.energy.2014.05.067}}.
\newline\urlprefix\url{https://doi.org/10.1016/j.energy.2014.05.067}

\bibitem{CHILD2016517}
M.~Child, C.~Breyer, \href{https://doi.org/10.1016/j.rser.2016.07.001}{{Vision
  and initial feasibility analysis of a recarbonised Finnish energy system for
  2050}}, Renewable and Sustainable Energy Reviews 66 (2016) 517 -- 536.
\newblock \href {http://dx.doi.org/10.1016/j.rser.2016.07.001}
  {\path{doi:10.1016/j.rser.2016.07.001}}.
\newline\urlprefix\url{https://doi.org/10.1016/j.rser.2016.07.001}

\bibitem{MATHIESEN2015139}
B.~Mathiesen, H.~Lund, D.~Connolly, H.~Wenzel, P.~Østergaard, B.~Möller,
  S.~Nielsen, I.~Ridjan, P.~Karnøe, K.~Sperling, F.~Hvelplund,
  \href{https://doi.org/10.1016/j.apenergy.2015.01.075}{{Smart Energy Systems
  for coherent 100\% renewable energy and transport solutions}}, Applied Energy
  145 (2015) 139 -- 154.
\newblock \href {http://dx.doi.org/10.1016/j.apenergy.2015.01.075}
  {\path{doi:10.1016/j.apenergy.2015.01.075}}.
\newline\urlprefix\url{https://doi.org/10.1016/j.apenergy.2015.01.075}

\bibitem{Palzer20141019}
A.~Palzer, H.-M. Henning, \href{https://doi.org/10.1016/j.rser.2013.11.032}{{A
  comprehensive model for the German electricity and heat sector in a future
  energy system with a dominant contribution from renewable energy technologies
  – Part II: Results}}, Renewable and Sustainable Energy Reviews 30 (2014)
  1019 -- 1034.
\newblock \href {http://dx.doi.org/10.1016/j.rser.2013.11.032}
  {\path{doi:10.1016/j.rser.2013.11.032}}.
\newline\urlprefix\url{https://doi.org/10.1016/j.rser.2013.11.032}

\bibitem{IJSEPM497}
D.~Connolly, B.~Mathiesen, \href{https://doi.org/10.5278/ijsepm.2014.1.2}{{A
  technical and economic analysis of one potential pathway to a 100\% renewable
  energy system}}, International Journal of Sustainable Energy Planning and
  Management 1~(0) (2014) 7--28.
\newblock \href {http://dx.doi.org/10.5278/ijsepm.2014.1.2}
  {\path{doi:10.5278/ijsepm.2014.1.2}}.
\newline\urlprefix\url{https://doi.org/10.5278/ijsepm.2014.1.2}

\bibitem{FERNANDES201451}
L.~Fernandes, P.~Ferreira,
  \href{https://doi.org/10.1016/j.energy.2014.02.098}{{Renewable energy
  scenarios in the Portuguese electricity system}}, Energy 69 (2014) 51 -- 57.
\newblock \href {http://dx.doi.org/10.1016/j.energy.2014.02.098}
  {\path{doi:10.1016/j.energy.2014.02.098}}.
\newline\urlprefix\url{https://doi.org/10.1016/j.energy.2014.02.098}

\bibitem{MOELLER201439}
C.~Moeller, J.~Meiss, B.~Mueller, M.~Hlusiak, C.~Breyer, M.~Kastner, J.~Twele,
  \href{https://doi.org/10.1016/j.renene.2014.06.042}{{Transforming the
  electricity generation of the Berlin–Brandenburg region, Germany}},
  Renewable Energy 72 (2014) 39 -- 50.
\newblock \href {http://dx.doi.org/10.1016/j.renene.2014.06.042}
  {\path{doi:10.1016/j.renene.2014.06.042}}.
\newline\urlprefix\url{https://doi.org/10.1016/j.renene.2014.06.042}

\bibitem{PIP:PIP2885}
C.~Breyer, D.~Bogdanov, A.~Gulagi, A.~Aghahosseini, L.~S. Barbosa, O.~Koskinen,
  M.~Barasa, U.~Caldera, S.~Afanasyeva, M.~Child, J.~Farfan, P.~Vainikka,
  \href{http://dx.doi.org/10.1002/pip.2885}{On the role of solar photovoltaics
  in global energy transition scenarios}, Progress in Photovoltaics: Research
  and Applications 25~(8) (2017) 727--745, pIP-16-176.R2.
\newblock \href {http://dx.doi.org/10.1002/pip.2885}
  {\path{doi:10.1002/pip.2885}}.
\newline\urlprefix\url{http://dx.doi.org/10.1002/pip.2885}

\bibitem{PIP:PIP2950}
C.~Breyer, D.~Bogdanov, A.~Aghahosseini, A.~Gulagi, M.~Child, A.~S. Oyewo,
  J.~Farfan, K.~Sadovskaia, P.~Vainikka,
  \href{http://dx.doi.org/10.1002/pip.2950}{Solar photovoltaics demand for the
  global energy transition in the power sector}, Progress in Photovoltaics:
  Research and Applications  n/a--n/aPIP-17-137.R1.
\newblock \href {http://dx.doi.org/10.1002/pip.2950}
  {\path{doi:10.1002/pip.2950}}.
\newline\urlprefix\url{http://dx.doi.org/10.1002/pip.2950}

\bibitem{Jacobson2017}
M.~Z. Jacobson, M.~A. Delucchi, Z.~A. Bauer, S.~C. Goodman, W.~E. Chapman,
  M.~A. Cameron, C.~Bozonnat, L.~Chobadi, H.~A. Clonts, P.~Enevoldsen, J.~R.
  Erwin, S.~N. Fobi, O.~K. Goldstrom, E.~M. Hennessy, J.~Liu, J.~Lo, C.~B.
  Meyer, S.~B. Morris, K.~R. Moy, P.~L. O'Neill, I.~Petkov, S.~Redfern,
  R.~Schucker, M.~A. Sontag, J.~Wang, E.~Weiner, A.~S. Yachanin,
  \href{https://doi.org/10.1016/j.joule.2017.07.005}{{100\% Clean and Renewable
  Wind, Water, and Sunlight All-Sector Energy Roadmaps for 139 Countries of the
  World}}, Joule 1 (2017) 1--14.
\newblock \href {http://dx.doi.org/10.1016/j.joule.2017.07.005}
  {\path{doi:10.1016/j.joule.2017.07.005}}.
\newline\urlprefix\url{https://doi.org/10.1016/j.joule.2017.07.005}

\bibitem{JACOBSON2018}
M.~Z. Jacobson, M.~A. Delucchi, M.~A. Cameron, B.~V. Mathiesen,
  \href{https://doi.org/10.1016/j.renene.2018.02.009}{Matching demand with
  supply at low cost in 139 countries among 20 world regions with 100\%
  intermittent wind, water, and sunlight (wws) for all purposes}, Renewable
  Energy\href {http://dx.doi.org/10.1016/j.renene.2018.02.009}
  {\path{doi:10.1016/j.renene.2018.02.009}}.
\newline\urlprefix\url{https://doi.org/10.1016/j.renene.2018.02.009}

\bibitem{su9020233}
A.~Gulagi, D.~Bogdanov, M.~Fasihi, C.~Breyer,
  \href{https://doi.org/10.3390/su9020233}{{Can Australia Power the
  Energy-Hungry Asia with Renewable Energy?}}, Sustainability 9~(2).
\newblock \href {http://dx.doi.org/10.3390/su9020233}
  {\path{doi:10.3390/su9020233}}.
\newline\urlprefix\url{https://doi.org/10.3390/su9020233}

\bibitem{en10050583}
A.~Gulagi, D.~Bogdanov, C.~Breyer, \href{https://doi.org/10.3390/en10050583}{{A
  Cost Optimized Fully Sustainable Power System for Southeast Asia and the
  Pacific Rim}}, Energies 10~(5).
\newblock \href {http://dx.doi.org/10.3390/en10050583}
  {\path{doi:10.3390/en10050583}}.
\newline\urlprefix\url{https://doi.org/10.3390/en10050583}

\bibitem{PLEMANN201719}
G.~Pleßmann, P.~Blechinger,
  \href{https://doi.org/10.1016/j.esr.2016.11.003}{{How to meet EU GHG emission
  reduction targets? A model based decarbonization pathway for Europe's
  electricity supply system until 2050}}, Energy Strategy Reviews 15 (2017) 19
  -- 32.
\newblock \href {http://dx.doi.org/10.1016/j.esr.2016.11.003}
  {\path{doi:10.1016/j.esr.2016.11.003}}.
\newline\urlprefix\url{https://doi.org/10.1016/j.esr.2016.11.003}

\bibitem{Schlachtberger2017}
D.~Schlachtberger, T.~Brown, S.~Schramm, M.~Greiner, The benefits of
  cooperation in a highly renewable {E}uropean electricity network, Energy 134
  (2017) 469 -- 481.
\newblock \href {http://dx.doi.org/10.1016/j.energy.2017.06.004}
  {\path{doi:10.1016/j.energy.2017.06.004}}.

\bibitem{Hoersch2017}
J.~H\"orsch, T.~Brown, \href{https://arxiv.org/abs/1705.07617}{The role of
  spatial scale in joint optimisations of generation and transmission for
  {E}uropean highly renewable scenarios}, in: Proceedings of 14th International
  Conference on the European Energy Market (EEM 2017), 2017.
\newblock \href {http://dx.doi.org/10.1109/EEM.2017.7982024}
  {\path{doi:10.1109/EEM.2017.7982024}}.
\newline\urlprefix\url{https://arxiv.org/abs/1705.07617}

\bibitem{GILS2017173}
H.~C. Gils, Y.~Scholz, T.~Pregger, D.~L. de~Tena, D.~Heide,
  \href{10.1016/j.energy.2017.01.115}{{Integrated modelling of variable
  renewable energy-based power supply in Europe}}, Energy 123 (2017) 173 --
  188.
\newblock \href
  {http://dx.doi.org/https://doi.org/10.1016/j.energy.2017.01.115}
  {\path{doi:https://doi.org/10.1016/j.energy.2017.01.115}}.
\newline\urlprefix\url{10.1016/j.energy.2017.01.115}

\bibitem{ERIKSEN2017913}
E.~H. Eriksen, L.~J. Schwenk-Nebbe, B.~Tranberg, T.~Brown, M.~Greiner,
  \href{https://doi.org/10.1016/j.energy.2017.05.170}{{Optimal heterogeneity in
  a simplified highly renewable European electricity system}}, Energy 133
  (2017) 913 -- 928.
\newblock \href {http://dx.doi.org/10.1016/j.energy.2017.05.170}
  {\path{doi:10.1016/j.energy.2017.05.170}}.
\newline\urlprefix\url{https://doi.org/10.1016/j.energy.2017.05.170}

\bibitem{CEBULLA2017211}
F.~Cebulla, T.~Naegler, M.~Pohl,
  \href{https://doi.org/10.1016/j.est.2017.10.004}{Electrical energy storage in
  highly renewable european energy systems: Capacity requirements, spatial
  distribution, and storage dispatch}, Journal of Energy Storage 14 (2017) 211
  -- 223.
\newblock \href {http://dx.doi.org/10.1016/j.est.2017.10.004}
  {\path{doi:10.1016/j.est.2017.10.004}}.
\newline\urlprefix\url{https://doi.org/10.1016/j.est.2017.10.004}

\bibitem{Brown2018}
T.~Brown, D.~Schlachtberger, A.~Kies, S.~Schramm, M.~Greiner,
  \href{https://arxiv.org/abs/1801.05290}{Synergies of sector coupling and
  transmission extension in a cost-optimised, highly renewable {E}uropean
  energy system}.
\newline\urlprefix\url{https://arxiv.org/abs/1801.05290}

\bibitem{PLEMANN2017}
G.~Pleßmann, P.~Blechinger,
  \href{https://doi.org/10.1016/j.energy.2017.03.076}{{Outlook on South-East
  European power system until 2050: Least-cost decarbonization pathway meeting
  EU mitigation targets}}, Energy\href
  {http://dx.doi.org/10.1016/j.energy.2017.03.076}
  {\path{doi:10.1016/j.energy.2017.03.076}}.
\newline\urlprefix\url{https://doi.org/10.1016/j.energy.2017.03.076}

\bibitem{10.1371/journal.pone.0173820}
L.~d. S. N.~S. Barbosa, D.~Bogdanov, P.~Vainikka, C.~Breyer,
  \href{https://doi.org/10.1371/journal.pone.0173820}{{Hydro, wind and solar
  power as a base for a 100\% renewable energy supply for South and Central
  America}}, PLOS ONE 12~(3) (2017) 1--28.
\newblock \href {http://dx.doi.org/10.1371/journal.pone.0173820}
  {\path{doi:10.1371/journal.pone.0173820}}.
\newline\urlprefix\url{https://doi.org/10.1371/journal.pone.0173820}

\bibitem{en10081171}
A.~Aghahosseini, D.~Bogdanov, C.~Breyer,
  \href{https://doi.org/10.3390/en10081171}{{A Techno-Economic Study of an
  Entirely Renewable Energy-Based Power Supply for North America for 2030
  Conditions}}, Energies 10~(8).
\newblock \href {http://dx.doi.org/10.3390/en10081171}
  {\path{doi:10.3390/en10081171}}.
\newline\urlprefix\url{https://doi.org/10.3390/en10081171}

\bibitem{GULAGI2017}
A.~Gulagi, D.~Bogdanov, C.~Breyer,
  \href{https://doi.org/10.1016/j.est.2017.11.012}{{The role of storage
  technologies in energy transition pathways towards achieving a fully
  sustainable energy system for India}}, Journal of Energy Storage\href
  {http://dx.doi.org/10.1016/j.est.2017.11.012}
  {\path{doi:10.1016/j.est.2017.11.012}}.
\newline\urlprefix\url{https://doi.org/10.1016/j.est.2017.11.012}

\bibitem{10.1371/journal.pone.0180611}
A.~Gulagi, P.~Choudhary, D.~Bogdanov, C.~Breyer,
  \href{https://doi.org/10.1371/journal.pone.0180611}{{Electricity system based
  on 100\% renewable energy for India and SAARC}}, PLOS ONE 12~(7) (2017)
  1--27.
\newblock \href {http://dx.doi.org/10.1371/journal.pone.0180611}
  {\path{doi:10.1371/journal.pone.0180611}}.
\newline\urlprefix\url{https://doi.org/10.1371/journal.pone.0180611}

\bibitem{BLAKERS2017471}
A.~Blakers, B.~Lu, M.~Stocks,
  \href{https://doi.org/10.1016/j.energy.2017.05.168}{{100\% renewable
  electricity in Australia}}, Energy 133 (2017) 471 -- 482.
\newblock \href {http://dx.doi.org/10.1016/j.energy.2017.05.168}
  {\path{doi:10.1016/j.energy.2017.05.168}}.
\newline\urlprefix\url{https://doi.org/10.1016/j.energy.2017.05.168}

\bibitem{LU2017663}
B.~Lu, A.~Blakers, M.~Stocks,
  \href{https://doi.org/10.1016/j.energy.2017.01.077}{{90–100\% renewable
  electricity for the South West Interconnected System of Western Australia}},
  Energy 122 (2017) 663 -- 674.
\newblock \href {http://dx.doi.org/10.1016/j.energy.2017.01.077}
  {\path{doi:10.1016/j.energy.2017.01.077}}.
\newline\urlprefix\url{https://doi.org/10.1016/j.energy.2017.01.077}

\bibitem{barbosa2016}
L.~de~Souza Noel Simas~Barbosa, J.~F. Orozco, D.~Bogdanov, P.~Vainikka,
  C.~Breyer, \href{https://doi.org/10.1016/j.egypro.2016.10.101}{{Hydropower
  and Power-to-gas Storage Options: The Brazilian Energy System Case}}, Energy
  Procedia 99 (2016) 89 -- 107, 10th International Renewable Energy Storage
  Conference, IRES 2016, 15-17 March 2016, Düsseldorf, Germany.
\newblock \href {http://dx.doi.org/10.1016/j.egypro.2016.10.101}
  {\path{doi:10.1016/j.egypro.2016.10.101}}.
\newline\urlprefix\url{https://doi.org/10.1016/j.egypro.2016.10.101}

\bibitem{Aghahosseini2017}
A.~Aghahosseini, D.~Bogdanov, N.~Ghorbani, C.~Breyer,
  \href{https://doi.org/10.1007/s13762-017-1373-4}{{Analysis of 100\% renewable
  energy for Iran in 2030: integrating solar PV, wind energy and storage}},
  International Journal of Environmental Science and Technology\href
  {http://dx.doi.org/10.1007/s13762-017-1373-4}
  {\path{doi:10.1007/s13762-017-1373-4}}.
\newline\urlprefix\url{https://doi.org/10.1007/s13762-017-1373-4}

\bibitem{SADIQA2018518}
A.~Sadiqa, A.~Gulagi, C.~Breyer,
  \href{https://doi.org/10.1016/j.energy.2018.01.027}{{Energy transition
  roadmap towards 100\% renewable energy and role of storage technologies for
  Pakistan by 2050}}, Energy 147 (2018) 518 -- 533.
\newblock \href {http://dx.doi.org/10.1016/j.energy.2018.01.027}
  {\path{doi:10.1016/j.energy.2018.01.027}}.
\newline\urlprefix\url{https://doi.org/10.1016/j.energy.2018.01.027}

\bibitem{Caldera2018}
U.~Caldera, D.~Bogdanov, S.~Afanasyeva, C.~Breyer,
  \href{https://doi.org/10.3390/w10010003}{{Role of Seawater Desalination in
  the Management of an Integrated Water and 100\% Renewable Energy Based Power
  Sector in Saudi Arabia}}, Water 10~(1).
\newblock \href {http://dx.doi.org/10.3390/w10010003}
  {\path{doi:10.3390/w10010003}}.
\newline\urlprefix\url{https://doi.org/10.3390/w10010003}

\bibitem{KILICKAPLAN2017218}
A.~Kilickaplan, D.~Bogdanov, O.~Peker, U.~Caldera, A.~Aghahosseini, C.~Breyer,
  \href{https://doi.org/10.1016/j.solener.2017.09.030}{{An energy transition
  pathway for Turkey to achieve 100\% renewable energy powered electricity,
  desalination and non-energetic industrial gas demand sectors by 2050}}, Solar
  Energy 158 (2017) 218 -- 235.
\newblock \href {http://dx.doi.org/10.1016/j.solener.2017.09.030}
  {\path{doi:10.1016/j.solener.2017.09.030}}.
\newline\urlprefix\url{https://doi.org/10.1016/j.solener.2017.09.030}

\bibitem{CHILD2017410}
M.~Child, C.~Breyer, D.~Bogdanov, H.-J. Fell,
  \href{https://doi.org/10.1016/j.egypro.2017.09.513}{{The role of storage
  technologies for the transition to a 100\% renewable energy system in
  Ukraine}}, Energy Procedia 135 (2017) 410 -- 423, 11th International
  Renewable Energy Storage Conference, IRES 2017, 14-16 March 2017,
  Düsseldorf, Germany.
\newblock \href {http://dx.doi.org/10.1016/j.egypro.2017.09.513}
  {\path{doi:10.1016/j.egypro.2017.09.513}}.
\newline\urlprefix\url{https://doi.org/10.1016/j.egypro.2017.09.513}

\bibitem{GILS2017342}
H.~C. Gils, S.~Simon,
  \href{https://doi.org/10.1016/j.apenergy.2016.12.023}{{Carbon neutral
  archipelago – 100\% renewable energy supply for the Canary Islands}},
  Applied Energy 188 (2017) 342 -- 355.
\newblock \href {http://dx.doi.org/10.1016/j.apenergy.2016.12.023}
  {\path{doi:10.1016/j.apenergy.2016.12.023}}.
\newline\urlprefix\url{https://doi.org/10.1016/j.apenergy.2016.12.023}

\bibitem{CHILD201749}
M.~Child, A.~Nordling, C.~Breyer,
  \href{https://doi.org/10.1016/j.enconman.2017.01.039}{{Scenarios for a
  sustainable energy system in the Åland Islands in 2030}}, Energy Conversion
  and Management 137 (2017) 49 -- 60.
\newblock \href {http://dx.doi.org/10.1016/j.enconman.2017.01.039}
  {\path{doi:10.1016/j.enconman.2017.01.039}}.
\newline\urlprefix\url{https://doi.org/10.1016/j.enconman.2017.01.039}

\bibitem{Martinot2007}
E.~Martinot, C.~Dienst, L.~Weiliang, C.~Qimin,
  \href{https://doi.org/10.1146/annurev.energy.32.080106.133554}{Renewable
  energy futures: Targets, scenarios, and pathways}, Annual Review of
  Environment and Resources 32~(1) (2007) 205--239.
\newblock \href {http://dx.doi.org/10.1146/annurev.energy.32.080106.133554}
  {\path{doi:10.1146/annurev.energy.32.080106.133554}}.
\newline\urlprefix\url{https://doi.org/10.1146/annurev.energy.32.080106.133554}

\bibitem{REN17}
{Renewables 2017 Global Status Report}, Tech. rep., REN21 (2017).

\bibitem{TRAINER20104107}
T.~Trainer, \href{https://doi.org/10.1016/j.enpol.2010.03.037}{Can renewables
  etc. solve the greenhouse problem? the negative case}, Energy Policy 38~(8)
  (2010) 4107 -- 4114.
\newblock \href {http://dx.doi.org/10.1016/j.enpol.2010.03.037}
  {\path{doi:10.1016/j.enpol.2010.03.037}}.
\newline\urlprefix\url{https://doi.org/10.1016/j.enpol.2010.03.037}

\bibitem{TRAINER2012476}
T.~Trainer, \href{https://doi.org/10.1016/j.enpol.2011.09.037}{A critique of
  jacobson and delucchi's proposals for a world renewable energy supply},
  Energy Policy 44 (2012) 476 -- 481.
\newblock \href {http://dx.doi.org/10.1016/j.enpol.2011.09.037}
  {\path{doi:10.1016/j.enpol.2011.09.037}}.
\newline\urlprefix\url{https://doi.org/10.1016/j.enpol.2011.09.037}

\bibitem{Clack2017}
C.~T.~M. Clack, S.~A. Qvist, J.~Apt, M.~Bazilian, A.~R. Brandt, K.~Caldeira,
  S.~J. Davis, V.~Diakov, M.~A. Handschy, P.~D.~H. Hines, P.~Jaramillo, D.~M.
  Kammen, J.~C.~S. Long, M.~G. Morgan, A.~Reed, V.~Sivaram, J.~Sweeney, G.~R.
  Tynan, D.~G. Victor, J.~P. Weyant, J.~F. Whitacre,
  \href{https://doi.org/10.1073/pnas.1610381114}{Evaluation of a proposal for
  reliable low-cost grid power with 100\% wind, water, and solar}, Proceedings
  of the National Academy of Sciences 114~(26) (2017) 6722--6727.
\newblock \href {http://dx.doi.org/10.1073/pnas.1610381114}
  {\path{doi:10.1073/pnas.1610381114}}.
\newline\urlprefix\url{https://doi.org/10.1073/pnas.1610381114}

\bibitem{TRAINER2013845}
T.~Trainer, \href{https://doi.org/10.1016/j.enpol.2013.09.027}{Can europe run
  on renewable energy? a negative case}, Energy Policy 63 (2013) 845 -- 850.
\newblock \href {http://dx.doi.org/10.1016/j.enpol.2013.09.027}
  {\path{doi:10.1016/j.enpol.2013.09.027}}.
\newline\urlprefix\url{https://doi.org/10.1016/j.enpol.2013.09.027}

\bibitem{Loftus2015}
P.~J. Loftus, A.~M. Cohen, J.~C.~S. Long, J.~D. Jenkins,
  \href{http://dx.doi.org/10.1002/wcc.324}{A critical review of global
  decarbonization scenarios: what do they tell us about feasibility?}, Wiley
  Interdisciplinary Reviews: Climate Change 6~(1) (2015) 93--112.
\newblock \href {http://dx.doi.org/10.1002/wcc.324}
  {\path{doi:10.1002/wcc.324}}.
\newline\urlprefix\url{http://dx.doi.org/10.1002/wcc.324}

\bibitem{Smil2010}
V.~Smil, Energy Transitions: History, Requirements, Prospects, Praeger
  Publishers, 2010.

\bibitem{DELUCCHI2012482}
M.~A. Delucchi, M.~Z. Jacobson,
  \href{https://doi.org/10.1016/j.enpol.2011.10.058}{Response to “a critique
  of jacobson and delucchi's proposals for a world renewable energy supply”
  by ted trainer}, Energy Policy 44 (2012) 482 -- 484.
\newblock \href {http://dx.doi.org/10.1016/j.enpol.2011.10.058}
  {\path{doi:10.1016/j.enpol.2011.10.058}}.
\newline\urlprefix\url{https://doi.org/10.1016/j.enpol.2011.10.058}

\bibitem{JACOBSON2013641}
M.~Z. Jacobson, M.~A. Delucchi,
  \href{https://doi.org/10.1016/j.enpol.2012.11.014}{Response to trainer's
  second commentary on a plan to power the world with wind, water, and solar
  power}, Energy Policy 57 (2013) 641 -- 643.
\newblock \href {http://dx.doi.org/10.1016/j.enpol.2012.11.014}
  {\path{doi:10.1016/j.enpol.2012.11.014}}.
\newline\urlprefix\url{https://doi.org/10.1016/j.enpol.2012.11.014}

\bibitem{Jacobson2017res}
M.~Z. Jacobson, M.~A. Delucchi, M.~A. Cameron, B.~A. Frew,
  \href{https://doi.org/10.1073/pnas.1708069114}{The united states can keep the
  grid stable at low cost with 100\% clean, renewable energy in all sectors
  despite inaccurate claims}, Proceedings of the National Academy of Sciences
  114~(26) (2017) E5021--E5023.
\newblock \href {http://dx.doi.org/10.1073/pnas.1708069114}
  {\path{doi:10.1073/pnas.1708069114}}.
\newline\urlprefix\url{https://doi.org/10.1073/pnas.1708069114}

\bibitem{burden}
B.~Heard, B.~Brook, T.~Wigley, C.~Bradshaw, Burden of proof: A comprehensive
  review of the feasibility of 100\% renewable-electricity systems, Renewable
  and Sustainable Energy Reviews 76 (2017) 1122 -- 1133.
\newblock \href {http://dx.doi.org/10.1016/j.rser.2017.03.114}
  {\path{doi:10.1016/j.rser.2017.03.114}}.

\bibitem{Brook2014}
B.~W. Brook, C.~J.~A. Bradshaw, \href{http://dx.doi.org/10.1111/cobi.12433}{Key
  role for nuclear energy in global biodiversity conservation}, Conservation
  Biology 29~(3) (2015) 702--712.
\newblock \href {http://dx.doi.org/10.1111/cobi.12433}
  {\path{doi:10.1111/cobi.12433}}.
\newline\urlprefix\url{http://dx.doi.org/10.1111/cobi.12433}

\bibitem{Heard2017}
B.~P. Heard, B.~W. Brook, \href{http://dx.doi.org/10.1002/app5.164}{{Closing
  the Cycle: How South Australia and Asia Can Benefit from Re-inventing Used
  Nuclear Fuel Management}}, Asia \& the Pacific Policy Studies 4~(1) (2017)
  166--175, aPPS-2016-0025.R2.
\newblock \href {http://dx.doi.org/10.1002/app5.164}
  {\path{doi:10.1002/app5.164}}.
\newline\urlprefix\url{http://dx.doi.org/10.1002/app5.164}

\bibitem{Brook2018}
B.~W. Brook, T.~Blees, T.~M.~L. Wigley, S.~Hong,
  \href{https://doi.org/10.3390/su10020302}{Silver buckshot or bullet: Is a
  future `energy mix' necessary?}, Sustainability 10~(2).
\newblock \href {http://dx.doi.org/10.3390/su10020302}
  {\path{doi:10.3390/su10020302}}.
\newline\urlprefix\url{https://doi.org/10.3390/su10020302}

\bibitem{IPCC2014-III-ATP}
L.~Clarke, K.~Jiang, K.~Akimoto, M.~Babiker, G.~Blanford, K.~Fisher-Vanden,
  J.-C. Hourcade, V.~Krey, E.~Kriegler, D.~M. A.~Löschel, S.~Paltsev, S.~Rose,
  P.~Shukla, M.~Tavoni, B.~van~der Zwaan, , D.~van Vuuren,
  \href{https://www.ipcc.ch/report/ar5/wg3/}{{Assessing Transformation
  Pathways}}, in: O.~Edenhofer, R.~Pichs-Madruga, Y.~Sokona, E.~Farahani,
  S.~Kadner, K.~Seyboth, A.~Adler, I.~Baum, S.~Brunner, P.~Eickemeier,
  B.~Kriemann, J.~Savolainen, S.~Schlömer, C.~von Stechow, T.~Zwickel, J.~Minx
  (Eds.), {Climate Change 2014: Mitigation of Climate Change. Contribution of
  Working Group III to the Fifth Assessment Report of the Intergovernmental
  Panel on Climate Change}, Cambridge University Press, Cambridge, United
  Kingdom and New York, NY, USA, 2014.
\newline\urlprefix\url{https://www.ipcc.ch/report/ar5/wg3/}

\bibitem{Kaldellis2010102}
J.~Kaldellis, \href{https://doi.org/10.1533/9781845699628.1.102}{4 -
  feasibility assessment for stand-alone and hybrid wind energy systems}, in:
  J.~Kaldellis (Ed.), Stand-Alone and Hybrid Wind Energy Systems, Woodhead
  Publishing Series in Energy, Woodhead Publishing, 2010, pp. 102 -- 161.
\newblock \href {http://dx.doi.org/10.1533/9781845699628.1.102}
  {\path{doi:10.1533/9781845699628.1.102}}.
\newline\urlprefix\url{https://doi.org/10.1533/9781845699628.1.102}

\bibitem{Bekele2010}
G.~Bekele, B.~Palm,
  \href{https://doi.org/10.1016/j.apenergy.2009.06.006}{Feasibility study for a
  standalone solar–wind-based hybrid energy system for application in
  ethiopia}, Applied Energy 87~(2) (2010) 487 -- 495.
\newblock \href {http://dx.doi.org/10.1016/j.apenergy.2009.06.006}
  {\path{doi:10.1016/j.apenergy.2009.06.006}}.
\newline\urlprefix\url{https://doi.org/10.1016/j.apenergy.2009.06.006}

\bibitem{CASTROSANTOS2016868}
L.~Castro-Santos, A.~Filgueira-Vizoso, L.~Carral-Couce, J.~Ángel
  Fraguela~Formoso,
  \href{https://doi.org/10.1016/j.energy.2016.06.135}{Economic feasibility of
  floating offshore wind farms}, Energy 112 (2016) 868 -- 882.
\newblock \href {http://dx.doi.org/10.1016/j.energy.2016.06.135}
  {\path{doi:10.1016/j.energy.2016.06.135}}.
\newline\urlprefix\url{https://doi.org/10.1016/j.energy.2016.06.135}

\bibitem{Rodrigues2016}
S.~Rodrigues, R.~Torabikalaki, F.~Faria, N.~Cafôfo, X.~Chen, A.~R. Ivaki,
  H.~Mata-Lima, F.~Morgado-Dias,
  \href{https://doi.org/10.1016/j.solener.2016.02.019}{Economic feasibility
  analysis of small scale pv systems in different countries}, Solar Energy 131
  (2016) 81 -- 95.
\newblock \href {http://dx.doi.org/10.1016/j.solener.2016.02.019}
  {\path{doi:10.1016/j.solener.2016.02.019}}.
\newline\urlprefix\url{https://doi.org/10.1016/j.solener.2016.02.019}

\bibitem{TSUPARI20171040}
E.~Tsupari, T.~Arponen, V.~Hankalin, J.~Kärki, S.~Kouri,
  \href{https://doi.org/10.1016/j.energy.2017.08.022}{Feasibility comparison of
  bioenergy and co2 capture and storage in a large combined heat, power and
  cooling system}, Energy 139 (2017) 1040 -- 1051.
\newblock \href {http://dx.doi.org/10.1016/j.energy.2017.08.022}
  {\path{doi:10.1016/j.energy.2017.08.022}}.
\newline\urlprefix\url{https://doi.org/10.1016/j.energy.2017.08.022}

\bibitem{RIDJAN201376}
I.~Ridjan, B.~V. Mathiesen, D.~Connolly, N.~Duić,
  \href{https://doi.org/10.1016/j.energy.2013.01.046}{The feasibility of
  synthetic fuels in renewable energy systems}, Energy 57 (2013) 76 -- 84.
\newblock \href {http://dx.doi.org/10.1016/j.energy.2013.01.046}
  {\path{doi:10.1016/j.energy.2013.01.046}}.
\newline\urlprefix\url{https://doi.org/10.1016/j.energy.2013.01.046}

\bibitem{macdonald2017}
A.~E. MacDonald, , C.~T.~M. Clack, A.~Alexander, A.~Dunbar, J.~Wilczak, Y.~Xie,
  \href{https://doi.org/10.1038/nclimate2921}{{Future cost-competitive
  electricity systems and their impact on US CO2 emissions}}, Nature Climate
  Change 6 (2017) 526--531.
\newblock \href {http://dx.doi.org/10.1038/nclimate2921}
  {\path{doi:10.1038/nclimate2921}}.
\newline\urlprefix\url{https://doi.org/10.1038/nclimate2921}

\bibitem{Rodriguez2013}
R.~Rodriguez, S.~Becker, G.~Andresen, D.~Heide, M.~Greiner, Transmission needs
  across a fully renewable {E}uropean power system, Renewable Energy 63 (2014)
  467--476.
\newblock \href {http://dx.doi.org/10.1016/j.renene.2013.10.005}
  {\path{doi:10.1016/j.renene.2013.10.005}}.

\bibitem{eurostat}
Eurostat, Eurostat, \url{http://ec.europa.eu/eurostat/}, {Online, retrieved
  June 2017}.

\bibitem{millstein2017}
D.~Millstein, R.~Wiser, M.~Bolinger, G.~Barbose,
  \href{https://doi.org/10.1038/nenergy.2017.134}{{The climate and air-quality
  benefits of wind and solar power in the United States}}, Nature Energy
  6~(17134).
\newblock \href {http://dx.doi.org/10.1038/nenergy.2017.134}
  {\path{doi:10.1038/nenergy.2017.134}}.
\newline\urlprefix\url{https://doi.org/10.1038/nenergy.2017.134}

\bibitem{IPCC2014-economic}
D.~Arent, R.~Tol, E.~Faust, J.~Hella, S.~Kumar, K.~Strzepek, F.~Tóth, ,
  D.~Yan, {Key economic sectors and services}, in: C.~Field, V.~Barros,
  D.~Dokken, K.~Mach, M.~Mastrandrea, T.~Bilir, M.~Chatterjee, K.~Ebi,
  Y.~Estrada, R.~Genova, B.~Girma, E.~Kissel, A.~Levy, S.~MacCracken,
  P.~Mastrandrea, L.~White (Eds.), {Climate Change 2014: Impacts, Adaptation,
  and Vulnerability. Part A: Global and Sectoral Aspects. Contribution of
  Working Group II to the Fifth Assessment Report of the Intergovernmental
  Panel on Climate Change}, Cambridge University Press, Cambridge, United
  Kingdom and New York, NY, USA, 2014, pp. 659--708,
  \url{https://www.ipcc.ch/pdf/assessment-report/ar5/wg2/WGIIAR5-Chap10_FINAL.pdf}.

\bibitem{TYNDP2016}
{European Network of Transmission System Operators for Electricity}, {Ten-Year
  Network Development Plan (TYNDP) 2016}, Tech. rep., ENTSO-E (2016).

\bibitem{er2015}
S.~Teske, S.~Sawyer, O.~Schäfer, T.~Pregger, S.~Simon, T.~Naegler, energy
  [r]evolution: a sustainable world energy outlook, 5th edition, Tech. rep.,
  Greenpeace International, Global Wind Energy Council (GWEC), SolarPowerEurope
  (2015).

\bibitem{weo2016}
{IEA}, \href{https://www.iea.org/weo/}{{World Energy Outlook 2016}}, Tech. rep.
  (2016).
\newline\urlprefix\url{https://www.iea.org/weo/}

\bibitem{Lund2014}
H.~Lund, {Renewable Energy Systems: A Smart Energy Systems Approach to the
  Choice and Modeling of 100\% Renewable Solutions, 2nd ed.}, Academic Press
  (Elsevier), 2014.

\bibitem{whitepaper2015}
{An electricity market for Germany's energy transition: White Paper}, Tech.
  rep., German Federal Ministry for Economic Affairs and Energy (BMWi),
  \url{https://www.bmwi.de/Redaktion/EN/Publikationen/whitepaper-electricity-market.html}
  (2015).

\bibitem{iwesgreen2015}
{Comments on the Green Paper `Ein Strommarkt für die Energiewende'}, Tech.
  rep., Fraunhofer-Institut für Windenergie und Energiesystemtechnik (IWES),
  status 16.06.2015, available at
  \url{http://www.bmwi.de/BMWi/Redaktion/PDF/Stellungnahmen-Gruenbuch/150226-fraunhofer-iwes-energiesystemtechnik,property=pdf,bereich=bmwi2012,sprache=de,rwb=true.pdf}
  (2015).

\bibitem{eurostatRE}
Eurostat, Eurostat renewable energy statistics explained,
  \url{http://ec.europa.eu/eurostat/statistics-explained/index.php/Renewable_energy_statistics},
  {Online, retrieved June 2017}.

\bibitem{WB2014}
{Sustainable Energy for All 2013-2014 : Global Tracking Framework}, Tech. rep.,
  World Bank and International Energy Agency (2014).
\newblock \href {http://dx.doi.org/10.1596/978-1-4648-0200-3}
  {\path{doi:10.1596/978-1-4648-0200-3}}.

\bibitem{Staffell2012}
I.~Staffell, D.~Brett, N.~Brandon, A.~Hawkes,
  \href{https://doi.org/10.1039/C2EE22653G}{A review of domestic heat pumps},
  Energy Environ. Sci. 5 (2012) 9291--9306.
\newblock \href {http://dx.doi.org/10.1039/C2EE22653G}
  {\path{doi:10.1039/C2EE22653G}}.
\newline\urlprefix\url{https://doi.org/10.1039/C2EE22653G}

\bibitem{MATHIESEN2012160}
B.~V. Mathiesen, H.~Lund, D.~Connolly,
  \href{https://doi.org/10.1016/j.energy.2012.07.063}{Limiting biomass
  consumption for heating in 100\% renewable energy systems}, Energy 48~(1)
  (2012) 160 -- 168, 6th Dubrovnik Conference on Sustainable Development of
  Energy Water and Environmental Systems, SDEWES 2011.
\newblock \href {http://dx.doi.org/10.1016/j.energy.2012.07.063}
  {\path{doi:10.1016/j.energy.2012.07.063}}.
\newline\urlprefix\url{https://doi.org/10.1016/j.energy.2012.07.063}

\bibitem{LUND20141}
H.~Lund, S.~Werner, R.~Wiltshire, S.~Svendsen, J.~E. Thorsen, F.~Hvelplund,
  B.~V. Mathiesen, \href{https://doi.org/10.1016/j.energy.2014.02.089}{4th
  generation district heating (4gdh)}, Energy 68 (2014) 1 -- 11.
\newblock \href {http://dx.doi.org/10.1016/j.energy.2014.02.089}
  {\path{doi:10.1016/j.energy.2014.02.089}}.
\newline\urlprefix\url{https://doi.org/10.1016/j.energy.2014.02.089}

\bibitem{CONNOLLY2014475}
D.~Connolly, H.~Lund, B.~Mathiesen, S.~Werner, B.~Möller, U.~Persson,
  T.~Boermans, D.~Trier, P.~Østergaard, S.~Nielsen,
  \href{https://doi.org/10.1016/j.enpol.2013.10.035}{{Heat Roadmap Europe:
  Combining district heating with heat savings to decarbonise the EU energy
  system}}, Energy Policy 65 (2014) 475 -- 489.
\newblock \href {http://dx.doi.org/10.1016/j.enpol.2013.10.035}
  {\path{doi:10.1016/j.enpol.2013.10.035}}.
\newline\urlprefix\url{https://doi.org/10.1016/j.enpol.2013.10.035}

\bibitem{eurostat-eb}
Energy balances,
  \url{http://ec.europa.eu/eurostat/web/energy/data/energy-balances} (2017).

\bibitem{vanVuuren2011}
D.~P. van Vuuren, J.~Edmonds, M.~Kainuma, K.~Riahi, A.~Thomson, K.~Hibbard,
  G.~C. Hurtt, T.~Kram, V.~Krey, J.-F. Lamarque, T.~Masui, M.~Meinshausen,
  N.~Nakicenovic, S.~J. Smith, S.~K. Rose, The representative concentration
  pathways: an overview, Climatic Change 109~(1) (2011) 5.
\newblock \href {http://dx.doi.org/10.1007/s10584-011-0148-z}
  {\path{doi:10.1007/s10584-011-0148-z}}.

\bibitem{Schellnhuber2016}
H.~J. Schellnhuber, S.~Rahmstorf, R.~Winkelmann,
  \href{http://dx.doi.org/10.1038/nclimate3013}{Why the right climate target
  was agreed in {P}aris}, Nature Climate Change 6 (2016) 649.
\newblock \href {http://dx.doi.org/10.1038/nclimate3013}
  {\path{doi:10.1038/nclimate3013}}.
\newline\urlprefix\url{http://dx.doi.org/10.1038/nclimate3013}

\bibitem{Creutzig2017}
F.~Creutzig, P.~Agoston, J.~C. Goldschmidt, G.~Luderer, G.~Nemet, R.~C.
  Pietzcker, \href{http://dx.doi.org/10.1038/nenergy.2017.140}{The
  underestimated potential of solar energy to mitigate climate change}, Nature
  Energy 2.
\newblock \href {http://dx.doi.org/10.1038/nenergy.2017.140}
  {\path{doi:10.1038/nenergy.2017.140}}.
\newline\urlprefix\url{http://dx.doi.org/10.1038/nenergy.2017.140}

\bibitem{IEESWV}
N.~Gerhardt, A.~Scholz, F.~Sandau, H.~Hahn,
  \href{{http://www.energiesystemtechnik.iwes.fraunhofer.de/de/projekte/suche/2015/interaktion_strom_waerme_verkehr.html}}{{Interaktion
  EE-Strom, Wärme und Verkehr}}, Tech. rep., Fraunhofer IWES (2015).
\newline\urlprefix\url{{http://www.energiesystemtechnik.iwes.fraunhofer.de/de/projekte/suche/2015/interaktion_strom_waerme_verkehr.html}}

\bibitem{EIAstudy2008}
{Carolyn Fischer and Evan Herrnstadt and Richard Morgenstern},
  \href{http://citeseerx.ist.psu.edu/viewdoc/download?doi=10.1.1.480.6720&rep=rep1&type=pdf}{{Understanding
  Errors in EIA Projections of Energy Demand}}, Tech. rep. (2008).
\newline\urlprefix\url{http://citeseerx.ist.psu.edu/viewdoc/download?doi=10.1.1.480.6720&rep=rep1&type=pdf}

\bibitem{martin2015}
C.~St.~Martin, J.~Lundquist, M.~Handschy,
  \href{http://stacks.iop.org/1748-9326/10/i=4/a=044004}{Variability of
  interconnected wind plants: correlation length and its dependence on
  variability time scale}, Environmental Research Letters 10 (2015) 044004.
\newblock \href {http://dx.doi.org/10.1088/1748-9326/10/4/044004}
  {\path{doi:10.1088/1748-9326/10/4/044004}}.
\newline\urlprefix\url{http://stacks.iop.org/1748-9326/10/i=4/a=044004}

\bibitem{norgard2004}
P.~B.~E. Nørgård, G.~Giebel, H.~Holttinen, L.~Söder, A.~Petterteig,
  \href{http://orbit.dtu.dk/files/7711241/ris_r_1443.pdf}{{WILMAR: Fluctuations
  and predictability of wind and hydropower: Deliverable 2.1}}, Tech. rep.,
  Forskningscenter Risoe (2004).
\newline\urlprefix\url{http://orbit.dtu.dk/files/7711241/ris_r_1443.pdf}

\bibitem{holttinen2011}
H.~Holttinen, J.~Kiviluoma, A.~Estanqueiro, E.~Gómez-Lázaro, B.~Rawn,
  J.~Dobschinski, P.~Meibom, E.~Lannoye, T.~Aigner, Y.~H. Wan, M.~Milligan,
  \href{https://www.ieawind.org/task_25/PDF/W1W/WIWAarhus_Task%2025%20Variability%20paper%20final.pdf}{{Variability
  of load and net load in case of large scale distributed wind power}}, in:
  10th Wind Integration Workshop, Aarhus, 2011.
\newline\urlprefix\url{https://www.ieawind.org/task_25/PDF/W1W/WIWAarhus_Task%2025%20Variability%20paper%20final.pdf}

\bibitem{ernst1999}
B.~Ernst, {Analysis of wind power ancillary services characteristics with
  German 250-MW wind data}, Tech. rep., National Renewable Energy Laboratory
  (1999).
\newblock \href {http://dx.doi.org/10.2172/752651} {\path{doi:10.2172/752651}}.

\bibitem{ernst}
B.~Ernst, T.~Bischof-Niemz, R.~van Heerden, C.~Mushwana, {Balancing wind energy
  within the South African power system}, in: 15th Wind Integration Workshop,
  Vienna, 2016.

\bibitem{DEANE2014152}
J.~Deane, G.~Drayton, B.~O. Gallach\'oir,
  \href{https://doi.org/10.1016/j.apenergy.2013.07.027}{The impact of
  sub-hourly modelling in power systems with significant levels of renewable
  generation}, Applied Energy 113 (2014) 152 -- 158.
\newblock \href {http://dx.doi.org/10.1016/j.apenergy.2013.07.027}
  {\path{doi:10.1016/j.apenergy.2013.07.027}}.
\newline\urlprefix\url{https://doi.org/10.1016/j.apenergy.2013.07.027}

\bibitem{6345631}
N.~Troy, D.~Flynn, M.~O'Malley,
  \href{https://doi.org/10.1109/PESGM.2012.6345631}{The importance of
  sub-hourly modeling with a high penetration of wind generation}, in: 2012
  IEEE Power and Energy Society General Meeting, 2012, pp. 1--6.
\newblock \href {http://dx.doi.org/10.1109/PESGM.2012.6345631}
  {\path{doi:10.1109/PESGM.2012.6345631}}.
\newline\urlprefix\url{https://doi.org/10.1109/PESGM.2012.6345631}

\bibitem{ODwyer2015}
C.~O'Dwyer, D.~Flynn, \href{https://doi.org/10.1109/TPWRS.2014.2356232}{Using
  energy storage to manage high net load variability at sub-hourly
  time-scales}, IEEE Transactions on Power Systems 30~(4) (2015) 2139--2148.
\newblock \href {http://dx.doi.org/10.1109/TPWRS.2014.2356232}
  {\path{doi:10.1109/TPWRS.2014.2356232}}.
\newline\urlprefix\url{https://doi.org/10.1109/TPWRS.2014.2356232}

\bibitem{DIMOULKAS2017}
I.~DIMOULKAS, M.~AMELIN, F.~LEVIHN,
  \href{https://doi.org/10.1007/s40565-017-0344-6}{District heating system
  operation in power systems with high share of wind power}, Journal of Modern
  Power Systems and Clean Energy 5~(6) (2017) 850--862.
\newblock \href {http://dx.doi.org/10.1007/s40565-017-0344-6}
  {\path{doi:10.1007/s40565-017-0344-6}}.
\newline\urlprefix\url{https://doi.org/10.1007/s40565-017-0344-6}

\bibitem{PFENNINGER20171}
S.~Pfenninger, \href{https://doi.org/10.1016/j.apenergy.2017.03.051}{{Dealing
  with multiple decades of hourly wind and PV time series in energy models: A
  comparison of methods to reduce time resolution and the planning implications
  of inter-annual variability}}, Applied Energy 197 (2017) 1 -- 13.
\newblock \href {http://dx.doi.org/10.1016/j.apenergy.2017.03.051}
  {\path{doi:10.1016/j.apenergy.2017.03.051}}.
\newline\urlprefix\url{https://doi.org/10.1016/j.apenergy.2017.03.051}

\bibitem{schroeder2013}
A.~Schr\"{o}der, F.~Kunz, J.~Meiss, R.~Mendelevitch, C.~von Hirschhausen,
  Current and prospective costs of electricity generation until 2050, Data
  Documentation, DIW~68, Deutsches Institut f\"{u}r Wirtschaftsforschung (DIW),
  Berlin, \url{http://hdl.handle.net/10419/80348}, accessed July 2016 (2013).

\bibitem{expertenkommission2015}
\href{https://www.bmwi.de/Redaktion/DE/Downloads/V/fuenfter-monitoring-bericht-energie-der-zukunft-stellungnahme.pdf}{{Stellungnahme
  zum fünften Monitoring-Bericht der Bundesregierung für das Berichtsjahr
  2015}}, Tech. rep., Expertenkommission zum Monitoring-Prozess `Energie der
  Zukunft' (2016).
\newline\urlprefix\url{https://www.bmwi.de/Redaktion/DE/Downloads/V/fuenfter-monitoring-bericht-energie-der-zukunft-stellungnahme.pdf}

\bibitem{IWESextreme}
N.~Gerhardt, D.~Böttger, T.~Trost, A.~Scholz, C.~Pape, A.-K. Gerlach,
  P.~Härtel, I.~Ganal,
  \href{{http://www.energieversorgung-elektromobilitaet.de/includes/reports/Auswertung_7Wetterjahre_95Prozent_FraunhoferIWES.pdf}}{{Analyse
  eines europäischen 95\%-Klimazielszenarios über mehrere Wetterjahre}},
  Tech. rep., Fraunhofer IWES (2017).
\newline\urlprefix\url{{http://www.energieversorgung-elektromobilitaet.de/includes/reports/Auswertung_7Wetterjahre_95Prozent_FraunhoferIWES.pdf}}

\bibitem{Roehrkasten2015}
S.~Roehrkasten, D.~Schaeuble, S.~Helgenberger,
  \href{https://doi.org/10.2312/iass.2015.023}{Secure and sustainable power
  generation in a water-constrained world}, IASS\href
  {http://dx.doi.org/10.2312/iass.2015.023} {\path{doi:10.2312/iass.2015.023}}.
\newline\urlprefix\url{https://doi.org/10.2312/iass.2015.023}

\bibitem{UEC13}
F.~Ueckerdt, L.~Hirth, G.~Luderer, O.~Edenhofer,
  \href{https://doi.org/10.1016/j.energy.2013.10.072}{{System LCOE: What are
  the costs of variable renewables?}}, Energy 63 (2013) 61--75.
\newblock \href {http://dx.doi.org/10.1016/j.energy.2013.10.072}
  {\path{doi:10.1016/j.energy.2013.10.072}}.
\newline\urlprefix\url{https://doi.org/10.1016/j.energy.2013.10.072}

\bibitem{HIR15}
L.~Hirth, F.~Ueckerdt, O.~Edenhofer,
  \href{https://doi.org/10.1016/j.renene.2014.08.065}{{Integration costs
  revisited -- An economic framework for wind and solar variability}},
  Renewable Energy 74 (2015) 925--939.
\newblock \href {http://dx.doi.org/10.1016/j.renene.2014.08.065}
  {\path{doi:10.1016/j.renene.2014.08.065}}.
\newline\urlprefix\url{https://doi.org/10.1016/j.renene.2014.08.065}

\bibitem{HESS2018874}
D.~Hess, M.~Wetzel, K.-K. Cao,
  \href{https://doi.org/10.1016/j.renene.2017.10.041}{Representing
  node-internal transmission and distribution grids in energy system models},
  Renewable Energy 119 (2018) 874 -- 890.
\newblock \href {http://dx.doi.org/10.1016/j.renene.2017.10.041}
  {\path{doi:10.1016/j.renene.2017.10.041}}.
\newline\urlprefix\url{https://doi.org/10.1016/j.renene.2017.10.041}

\bibitem{ICNERA}
{Integration of Renewable Energy in Europe}, Tech. rep., Imperial College,
  NERA, DNV GL,
  \url{https://ec.europa.eu/energy/sites/ener/files/documents/201406_report_renewables_integration_europe.pdf}
  (2014).

\bibitem{RLP}
T.~Ackermann, N.~Martensen, T.~Brown, et~al.,
  \href{https://www.oeko.de/oekodoc/1885/2014-008-de.pdf}{Verteilnetzstudie
  Rheinland-Pfalz}, 2014.
\newline\urlprefix\url{https://www.oeko.de/oekodoc/1885/2014-008-de.pdf}

\bibitem{Brown}
T.~Brown, P.~Schierhorn, E.~Tr\"oster, T.~Ackermann, Optimising the {European}
  transmission system for 77\% renewable electricity by 2030, IET Renewable
  Power Generation 10~(1) (2016) 3--9.
\newblock \href {http://dx.doi.org/10.1049/iet-rpg.2015.0135}
  {\path{doi:10.1049/iet-rpg.2015.0135}}.

\bibitem{energynautics}
E.~Tr{\"o}ster, R.~Kuwahata, T.~Ackermann, European grid study 2030/2050, Tech.
  rep., {E}nergynautics {G}mb{H} (2011).

\bibitem{DENA12}
{Deutsche Energie-Agentur},
  \href{http://www.dena.de/publikationen/energiesysteme}{{DENA-Verteilnetzstudie}}
  (2012).
\newline\urlprefix\url{http://www.dena.de/publikationen/energiesysteme}

\bibitem{Becker2012}
S.~Becker, R.~Rodriguez, G.~Andresen, S.~Schramm, M.~Greiner, Transmission grid
  extensions during the build-up of a fully renewable pan-{E}uropean
  electricity supply, Energy 64 (2014) 404--418.
\newblock \href {http://dx.doi.org/10.1016/j.energy.2013.10.010}
  {\path{doi:10.1016/j.energy.2013.10.010}}.

\bibitem{ACER2016}
{ACER},
  \href{http://www.acer.europa.eu/Official_documents/Acts_of_the_Agency/Publication/ACER%20Market%20Monitoring%20Report%202016%20-%20ELECTRICITY.pdf}{{ACER
  Market Monitoring Report 2016}}, Tech. rep. (2017).
\newline\urlprefix\url{http://www.acer.europa.eu/Official_documents/Acts_of_the_Agency/Publication/ACER%20Market%20Monitoring%20Report%202016%20-%20ELECTRICITY.pdf}

\bibitem{amprionbiblis}
Amprion, {Pressemitteilung: Generator wird zum Motor},
  \url{https://www.amprion.net/Presse/Presse-Detailseite_2667.html} (2012).

\bibitem{DNVSys12}
W.~Uijlings, {System Service Provision: An independent view on the likely costs
  incurred by potential System Service Providers in delivering additional and
  enhanced System Services}, Tech. rep., DNV KEMA (2012).

\bibitem{DENA16}
{Deutsche Energie-Agentur},
  \href{https://shop.dena.de/fileadmin/denashop/media/Downloads_Dateien/esd/9142_Studie_Momentanreserve_2030.pdf}{{Momentanreserve
  2030: Bedarf und Erbringung von Momentanreserve 2030}} (2016).
\newline\urlprefix\url{https://shop.dena.de/fileadmin/denashop/media/Downloads_Dateien/esd/9142_Studie_Momentanreserve_2030.pdf}

\bibitem{tennetgermany}
TenneT,
  \href{https://www.tennet.eu/de/news/news/tennet-gestaltet-die-energiewende-in-bayern-360-tonnen-schwerer-generator-erreicht-umspannwerk-berg/}{{TenneT
  gestaltet die Energiewende in Bayern: 360 Tonnen schwerer Generator erreicht
  Umspannwerk Bergrheinfeld West}} (2015).
\newline\urlprefix\url{https://www.tennet.eu/de/news/news/tennet-gestaltet-die-energiewende-in-bayern-360-tonnen-schwerer-generator-erreicht-umspannwerk-berg/}

\bibitem{siemenssc}
Siemens,
  \href{https://w3.usa.siemens.com/smartgrid/us/en/events/Documents/IEEE%202016/TS_Synchronous_Condenser.pdf}{{The
  stable way: Synchronous condenser solutions}} (2014).
\newline\urlprefix\url{https://w3.usa.siemens.com/smartgrid/us/en/events/Documents/IEEE%202016/TS_Synchronous_Condenser.pdf}

\bibitem{Tasmania2016}
{Managing a High Penetration of Renewables - A Tasmanian Case Study}, Tech.
  rep., Hydro Tasmania and TasNetworks (Aug 2016).

\bibitem{Orths2016}
A.~Orths, P.~B. Eriksen,
  \href{http://dx.doi.org/10.23723/1301:2016-5/17783}{The future has come: the
  100\% res driven power system is reality}, Revue de l’Electricité et de
  l’Electronique 5.
\newblock \href {http://dx.doi.org/10.23723/1301:2016-5/17783}
  {\path{doi:10.23723/1301:2016-5/17783}}.
\newline\urlprefix\url{http://dx.doi.org/10.23723/1301:2016-5/17783}

\bibitem{Denmark2018}
\href{https://stateofgreen.com/en/profiles/state-of-green/news/danish-electrical-grid-independent-of-centralised-power-plants-for-41-days}{{Danish
  Electrical Grid Independent of Centralised Power Plants for 41 Days}} (2018).
\newline\urlprefix\url{https://stateofgreen.com/en/profiles/state-of-green/news/danish-electrical-grid-independent-of-centralised-power-plants-for-41-days}

\bibitem{miller2015}
N.~W. Miller, Keeping it together: Transient stability in a world of wind and
  solar generation, IEEE Power and Energy Magazine 13~(6) (2015) 31--39.
\newblock \href {http://dx.doi.org/10.1109/MPE.2015.2461332}
  {\path{doi:10.1109/MPE.2015.2461332}}.

\bibitem{wwsis3a}
N.~W. Miller, B.~Leonardi, R.~D'Aquila, K.~Clark, Tech. rep., NREL (Nov 2015).
\newblock \href{http://www.nrel.gov/docs/fy16osti/64822.pdf}{[link]}.
\newline\urlprefix\url{http://www.nrel.gov/docs/fy16osti/64822.pdf}

\bibitem{urdal2015}
H.~Urdal, R.~Ierna, J.~Zhu, C.~Ivanov, A.~Dahresobh, D.~Rostom, System strength
  considerations in a converter dominated power system, IET Renewable Power
  Generation 9 (2015) 10--17(7).
\newblock \href {http://dx.doi.org/10.1049/iet-rpg.2014.0199}
  {\path{doi:10.1049/iet-rpg.2014.0199}}.

\bibitem{Kueck200627}
J.~Kueck, B.~Kirby, T.~Rizy, F.~Li, N.~Fall, Reactive power from distributed
  energy, The Electricity Journal 19~(10) (2006) 27 -- 38.
\newblock \href {http://dx.doi.org/10.1016/j.tej.2006.10.007}
  {\path{doi:10.1016/j.tej.2006.10.007}}.

\bibitem{Igbinovia2016}
F.~O. Igbinovia, G.~Fandi, Z.~Müller, J.~Švec, J.~Tlusty, Cost implication
  and reactive power generating potential of the synchronous condenser, in:
  2016 2nd International Conference on Intelligent Green Building and Smart
  Grid (IGBSG), 2016, pp. 1--6.
\newblock \href {http://dx.doi.org/10.1109/IGBSG.2016.7539450}
  {\path{doi:10.1109/IGBSG.2016.7539450}}.

\bibitem{OTHINA}
{Zukünftige Bereitstellung von Blindleistung und anderen Maßnahmen für die
  Netzsicherheit}, Tech. rep., INA, OTH (2016).

\bibitem{gesc}
G.~E. Company, {Synchronous Condensers for Transmission Systems},
  \url{http://www.ercot.com/content/meetings/rpg/keydocs/2013/1115/GE_Condensers_Ercot_presentation_(11-15-13).pdf}
  (2013).

\bibitem{Kroposki2017}
B.~Kroposki, B.~Johnson, Y.~Zhang, V.~Gevorgian, P.~Denholm, B.~M. Hodge,
  B.~Hannegan, Achieving a 100\% renewable grid: Operating electric power
  systems with extremely high levels of variable renewable energy, IEEE Power
  and Energy Magazine 15~(2) (2017) 61--73.
\newblock \href {http://dx.doi.org/10.1109/MPE.2016.2637122}
  {\path{doi:10.1109/MPE.2016.2637122}}.

\bibitem{Milligan2015}
M.~Milligan, B.~Frew, B.~Kirby, M.~Schuerger, K.~Clark, D.~Lew, P.~Denholm,
  B.~Zavadil, M.~O'Malley, B.~Tsuchida,
  \href{https://doi.org/10.1109/MPE.2015.2462311}{Alternatives no more: Wind
  and solar power are mainstays of a clean, reliable, affordable grid}, IEEE
  Power and Energy Magazine 13~(6) (2015) 78--87.
\newblock \href {http://dx.doi.org/10.1109/MPE.2015.2462311}
  {\path{doi:10.1109/MPE.2015.2462311}}.
\newline\urlprefix\url{https://doi.org/10.1109/MPE.2015.2462311}

\bibitem{IRENA2016}
T.~Ackermann, N.~Martensen, T.~Brown, P.-P. Schierhorn, F.~Boshell, F.~Gafaro,
  M.~Ayuso,
  \href{http://www.irena.org/DocumentDownloads/Publications/IRENA_Grid_Codes_2016.pdf}{{Scaling
  Up Variable Renewable Power: The Role of Grid Codes}}, Tech. rep.,
  International Renewable Energy Agency (IRENA) (2016).
\newline\urlprefix\url{http://www.irena.org/DocumentDownloads/Publications/IRENA_Grid_Codes_2016.pdf}

\bibitem{macdowell2015}
J.~MacDowell, S.~Dutta, M.~Richwine, S.~Achilles, N.~Miller,
  \href{https://doi.org/10.1109/MPE.2015.2461331}{{Serving the Future: Advanced
  Wind Generation Technology Supports Ancillary Services}}, IEEE Power and
  Energy Magazine 13~(6) (2015) 22--30.
\newblock \href {http://dx.doi.org/10.1109/MPE.2015.2461331}
  {\path{doi:10.1109/MPE.2015.2461331}}.
\newline\urlprefix\url{https://doi.org/10.1109/MPE.2015.2461331}

\bibitem{rocof2}
\href{http://www.eirgridgroup.com/site-files/library/EirGrid/RoCoF-Alternative-Solutions-Project-Phase-2-Report-Final.pdf}{{RoCoF
  Alternative \& Complementary Solutions Project: Phase 2 Study Report}}, Tech.
  rep., EirGrid, SONI (2016).
\newline\urlprefix\url{http://www.eirgridgroup.com/site-files/library/EirGrid/RoCoF-Alternative-Solutions-Project-Phase-2-Report-Final.pdf}

\bibitem{strathprints58052}
R.~Ierna, J.~Zhu, A.~J. Roscoe, M.~Yu, A.~Dysko, C.~D. Booth, H.~Urdal,
  \href{http://strathprints.strath.ac.uk/58052/}{Effects of {VSM} convertor
  control on penetration limits of non-synchronous generation in the {GB} power
  system}, in: 15th Wind Integration Workshop, 2016, this paper was presented
  at the 15th Wind Integration Workshop and published in the workshop's
  proceedings.
\newline\urlprefix\url{http://strathprints.strath.ac.uk/58052/}

\bibitem{strathprints58053}
A.~J. Roscoe, M.~Yu, R.~Ierna, J.~Zhu, A.~Dy{\'s}ko, H.~Urdal, C.~Booth,
  \href{http://strathprints.strath.ac.uk/58053/}{A {VSM} (virtual synchronous
  machine) convertor control model suitable for {RMS} studies for resolving
  system operator/owner challenges}, in: 15th Wind Integration Workshop, 2016,
  this paper was presented at the 15th Wind Integration Workshop and published
  in the workshop's proceedings.
\newline\urlprefix\url{http://strathprints.strath.ac.uk/58053/}

\bibitem{chen2011}
Y.~Chen, R.~Hesse, D.~Turschner, H.~P. Beck,
  \href{https://doi.org/10.1109/PowerEng.2011.6036498}{Improving the grid power
  quality using virtual synchronous machines}, in: 2011 International
  Conference on Power Engineering, Energy and Electrical Drives, 2011, pp.
  1--6.
\newblock \href {http://dx.doi.org/10.1109/PowerEng.2011.6036498}
  {\path{doi:10.1109/PowerEng.2011.6036498}}.
\newline\urlprefix\url{https://doi.org/10.1109/PowerEng.2011.6036498}

\bibitem{karapanos2011}
V.~Karapanos, S.~de~Haan, K.~Zwetsloot,
  \href{https://doi.org/10.1109/IECON.2011.6119919}{Real time simulation of a
  power system with {VSG} hardware in the loop}, in: IECON 2011 - 37th Annual
  Conference of the IEEE Industrial Electronics Society, 2011, pp. 3748--3754.
\newblock \href {http://dx.doi.org/10.1109/IECON.2011.6119919}
  {\path{doi:10.1109/IECON.2011.6119919}}.
\newline\urlprefix\url{https://doi.org/10.1109/IECON.2011.6119919}

\bibitem{zhong2011}
Q.~C. Zhong, G.~Weiss,
  \href{https://doi.org/10.1109/TIE.2010.2048839}{Synchronverters: Inverters
  that mimic synchronous generators}, IEEE Transactions on Industrial
  Electronics 58~(4) (2011) 1259--1267.
\newblock \href {http://dx.doi.org/10.1109/TIE.2010.2048839}
  {\path{doi:10.1109/TIE.2010.2048839}}.
\newline\urlprefix\url{https://doi.org/10.1109/TIE.2010.2048839}

\bibitem{torres2013}
M.~Torres, L.~Lopes, \href{https://doi.org/10.4236/epe.2013.52A005}{{Virtual
  Synchronous Generator: A Control Strategy to Improve Dynamic Frequency
  Control in Autonomous Power Systems}}, Energy and Power Engineering 5~(2A)
  (2013) 32--38.
\newblock \href {http://dx.doi.org/10.4236/epe.2013.52A005}
  {\path{doi:10.4236/epe.2013.52A005}}.
\newline\urlprefix\url{https://doi.org/10.4236/epe.2013.52A005}

\bibitem{bevrani2014}
H.~Bevrani, T.~Ise, Y.~Miura,
  \href{https://doi.org/10.1016/j.ijepes.2013.07.009}{Virtual synchronous
  generators: A survey and new perspectives}, International Journal of
  Electrical Power \& Energy Systems 54 (2014) 244 -- 254.
\newblock \href {http://dx.doi.org/10.1016/j.ijepes.2013.07.009}
  {\path{doi:10.1016/j.ijepes.2013.07.009}}.
\newline\urlprefix\url{https://doi.org/10.1016/j.ijepes.2013.07.009}

\bibitem{jouini2016}
T.~Jouini, C.~Arghir, F.~Dörfler,
  \href{https://doi.org/10.1016/j.ifacol.2016.10.395}{{Grid-Friendly Matching
  of Synchronous Machines by Tapping into the DC Storage}}, IFAC-PapersOnLine
  49~(22) (2016) 192 -- 197, 6th IFAC Workshop on Distributed Estimation and
  Control in Networked Systems NECSYS 2016.
\newblock \href {http://dx.doi.org/10.1016/j.ifacol.2016.10.395}
  {\path{doi:10.1016/j.ifacol.2016.10.395}}.
\newline\urlprefix\url{https://doi.org/10.1016/j.ifacol.2016.10.395}

\bibitem{sinha2017}
M.~Sinha, F.~Dörfler, B.~B. Johnson, S.~V. Dhople,
  \href{https://doi.org/10.1109/TCNS.2015.2503558}{{Uncovering Droop Control
  Laws Embedded Within the Nonlinear Dynamics of Van der Pol Oscillators}},
  IEEE Transactions on Control of Network Systems 4~(2) (2017) 347--358.
\newblock \href {http://dx.doi.org/10.1109/TCNS.2015.2503558}
  {\path{doi:10.1109/TCNS.2015.2503558}}.
\newline\urlprefix\url{https://doi.org/10.1109/TCNS.2015.2503558}

\bibitem{IRENA2015}
{Technology Brief: Renewable Energy Integration in Power Grids}, Tech. rep.,
  IRENA, IEA-ETSAP (2015).

\bibitem{DENASS}
{dena-Studie Systemdienstleistungen 2030}, Tech. rep., DENA, online at
  \url{https://www.dena.de/themen-projekte/projekte/energiesysteme/dena-studie-systemdienstleistungen-2030/}
  (2014).

\bibitem{KWK2}
Kombikraftwerk 2: Abschlussbericht, Tech. rep., Fraunhofer IWES and others
  (August 2014).

\bibitem{BNetzA2016}
{BNetzA}, {BKartA},
  \href{https://www.bundesnetzagentur.de/DE/Sachgebiete/ElektrizitaetundGas/Unternehmen_Institutionen/DatenaustauschundMonitoring/Monitoring/Monitoringberichte/Monitoring_Berichte_node.html}{Monitoringbericht
  2016} (2016).
\newline\urlprefix\url{https://www.bundesnetzagentur.de/DE/Sachgebiete/ElektrizitaetundGas/Unternehmen_Institutionen/DatenaustauschundMonitoring/Monitoring/Monitoringberichte/Monitoring_Berichte_node.html}

\bibitem{wemag-black}
WEMAG,
  \href{https://www.wemag.com/aktuelles-presse/wemag-batteriespeicher-testet-erfolgreich-schwarzstart-nach-blackout}{{WEMAG-Batteriespeicher
  testet erfolgreich Schwarzstart nach Blackout}} (2017).
\newline\urlprefix\url{https://www.wemag.com/aktuelles-presse/wemag-batteriespeicher-testet-erfolgreich-schwarzstart-nach-blackout}

\bibitem{california-black}
E.~S. News,
  \href{https://www.energy-storage.news/news/california-batterys-black-start-capability-hailed-as-major-accomplishment-i}{California
  battery's black start capability hailed as `major accomplishment in the
  energy industry'} (2017).
\newline\urlprefix\url{https://www.energy-storage.news/news/california-batterys-black-start-capability-hailed-as-major-accomplishment-i}

\bibitem{uranium2016}
{Uranium 2016: Resources, Production and Demand}, Tech. rep., Nuclear Energy
  Agency and International Atomic Energy Agency (2016).

\bibitem{abbott2011}
D.~Abbott, \href{https://doi.org/10.1109/JPROC.2011.2161806}{{Is Nuclear Power
  Globally Scalable?}}, Proceedings of the IEEE 99~(10) (2011) 1611--1617.
\newblock \href {http://dx.doi.org/10.1109/JPROC.2011.2161806}
  {\path{doi:10.1109/JPROC.2011.2161806}}.
\newline\urlprefix\url{https://doi.org/10.1109/JPROC.2011.2161806}

\bibitem{iea2016}
\href{http://www.iea.org/publications/freepublications/publication/KeyWorld2016.pdf}{{Key
  world energy statistics 2016}}, Tech. rep., International Energy Agency
  (2016).
\newline\urlprefix\url{http://www.iea.org/publications/freepublications/publication/KeyWorld2016.pdf}

\bibitem{Perez2015}
R.~Perez, M.~Perez,
  \href{https://www.iea-shc.org/data/sites/1/publications/2015-11-A-Fundamental-Look-at-Supply-Side-Energy-Reserves-for-the-Planet.pdf}{A
  fundamental look at supply side energy reserves for the planet}, IEA SHC.
\newline\urlprefix\url{https://www.iea-shc.org/data/sites/1/publications/2015-11-A-Fundamental-Look-at-Supply-Side-Energy-Reserves-for-the-Planet.pdf}

\bibitem{VERBRUGGEN20084036}
A.~Verbruggen, \href{https://doi.org/10.1016/j.enpol.2008.06.024}{Renewable and
  nuclear power: A common future?}, Energy Policy 36~(11) (2008) 4036 -- 4047,
  transition towards Sustainable Energy Systems.
\newblock \href {http://dx.doi.org/10.1016/j.enpol.2008.06.024}
  {\path{doi:10.1016/j.enpol.2008.06.024}}.
\newline\urlprefix\url{https://doi.org/10.1016/j.enpol.2008.06.024}

\bibitem{HAMACHER2013657}
T.~Hamacher, M.~Huber, J.~Dorfner, K.~Schaber, A.~M. Bradshaw,
  \href{https://doi.org/10.1016/j.fusengdes.2013.01.074}{Nuclear fusion and
  renewable energy forms: Are they compatible?}, Fusion Engineering and Design
  88~(6) (2013) 657 -- 660, proceedings of the 27th Symposium On Fusion
  Technology (SOFT-27); Liège, Belgium, September 24-28, 2012.
\newblock \href {http://dx.doi.org/10.1016/j.fusengdes.2013.01.074}
  {\path{doi:10.1016/j.fusengdes.2013.01.074}}.
\newline\urlprefix\url{https://doi.org/10.1016/j.fusengdes.2013.01.074}

\bibitem{en10040578}
A.~David, B.~V. Mathiesen, H.~Averfalk, S.~Werner, H.~Lund,
  \href{https://doi.org/10.3390/en10040578}{{Heat Roadmap Europe: Large-Scale
  Electric Heat Pumps in District Heating Systems}}, Energies 10~(4).
\newblock \href {http://dx.doi.org/10.3390/en10040578}
  {\path{doi:10.3390/en10040578}}.
\newline\urlprefix\url{https://doi.org/10.3390/en10040578}

\bibitem{Tesla100MW2017}
\href{https://www.theguardian.com/australia-news/2017/dec/01/south-australia-turns-on-teslas-100mw-battery-history-in-the-making}{{South
  Australia turns on Tesla's 100MW battery}} (Dec 2017).
\newline\urlprefix\url{https://www.theguardian.com/australia-news/2017/dec/01/south-australia-turns-on-teslas-100mw-battery-history-in-the-making}

\bibitem{EIA2018}
\href{https://www.eia.gov/todayinenergy/detail.php?id=34432}{{Batteries perform
  many different functions on the power grid}} (Jan 2018).
\newline\urlprefix\url{https://www.eia.gov/todayinenergy/detail.php?id=34432}

\bibitem{INL2015}
H.~Gougar, R.~Bari, T.~Kim, T.~E. Sowinski, A.~Worrall,
  \href{https://inldigitallibrary.inl.gov/sites/sti/sti/6721146.pdf}{{Assessment
  of the Technical Maturity of Generation IV Concepts for Test or Demonstration
  Reactor Applications}}, Tech. rep., Idaho National Laboratory (2015).
\newline\urlprefix\url{https://inldigitallibrary.inl.gov/sites/sti/sti/6721146.pdf}

\bibitem{cochran2010}
T.~B. Cochran, H.~A. Feiveson, Z.~Mian, M.~V. Ramana, M.~Schneider, F.~N. von
  Hippel, \href{https://doi.org/10.2968/066003007}{It's time to give up on
  breeder reactors}, Bulletin of the Atomic Scientists 66~(3) (2010) 50--56.
\newblock \href {http://dx.doi.org/10.2968/066003007}
  {\path{doi:10.2968/066003007}}.
\newline\urlprefix\url{https://doi.org/10.2968/066003007}

\bibitem{slowdeath}
J.~Green,
  \href{https://www.wiseinternational.org/nuclear-monitor/831/slow-death-fast-reactors}{The
  slow death of fast reactors}, Nuclear Monitor~(4587).
\newline\urlprefix\url{https://www.wiseinternational.org/nuclear-monitor/831/slow-death-fast-reactors}

\bibitem{EFDA}
A roadmap to the realisation of fusion energy, Tech. rep., EFDA (2012).

\bibitem{IPCC2014-III-SPM}
\href{https://www.ipcc.ch/report/ar5/wg3/}{{Summary for Policy Makers}}, in:
  O.~Edenhofer, R.~Pichs-Madruga, Y.~Sokona, E.~Farahani, S.~Kadner,
  K.~Seyboth, A.~Adler, I.~Baum, S.~Brunner, P.~Eickemeier, B.~Kriemann,
  J.~Savolainen, S.~Schlömer, C.~von Stechow, T.~Zwickel, J.~Minx (Eds.),
  {Climate Change 2014: Mitigation of Climate Change. Contribution of Working
  Group III to the Fifth Assessment Report of the Intergovernmental Panel on
  Climate Change}, Cambridge University Press, Cambridge, United Kingdom and
  New York, NY, USA, 2014.
\newline\urlprefix\url{https://www.ipcc.ch/report/ar5/wg3/}

\bibitem{Nykvist2015}
B.~Nykvist, M.~Nilsson, \href{https://doi.org/10.1038/nclimate2564}{Rapidly
  falling costs of battery packs for electric vehicles}, Nature Clim. Change 7
  (2015) 329--332.
\newblock \href {http://dx.doi.org/10.1038/nclimate2564}
  {\path{doi:10.1038/nclimate2564}}.
\newline\urlprefix\url{https://doi.org/10.1038/nclimate2564}

\bibitem{schmidt2017}
O.~Schmidt, A.~Hawkes, A.~Gambhir, I.~Staffell,
  \href{https://doi.org/10.1038/nenergy.2017.110}{{The future cost of
  electrical energy storage based on experience rates}}, Nature Energy
  2~(17110).
\newblock \href {http://dx.doi.org/10.1038/nenergy.2017.110}
  {\path{doi:10.1038/nenergy.2017.110}}.
\newline\urlprefix\url{https://doi.org/10.1038/nenergy.2017.110}

\bibitem{kittner2017}
N.~Kittner, F.~Lill, D.~M. Kammen,
  \href{https://doi.org/10.1038/nenergy.2017.125}{{Energy storage deployment
  and innovation for the clean energy transition}}, Nature Energy 2~(17125).
\newblock \href {http://dx.doi.org/10.1038/nenergy.2017.125}
  {\path{doi:10.1038/nenergy.2017.125}}.
\newline\urlprefix\url{https://doi.org/10.1038/nenergy.2017.125}

\bibitem{AFANASYEVA2016157}
S.~Afanasyeva, C.~Breyer, M.~Engelhard,
  \href{https://doi.org/10.1016/j.egypro.2016.10.107}{Impact of battery cost on
  the economics of hybrid photovoltaic power plants}, Energy Procedia 99 (2016)
  157 -- 173, 10th International Renewable Energy Storage Conference, IRES
  2016, 15-17 March 2016, Düsseldorf, Germany.
\newblock \href {http://dx.doi.org/10.1016/j.egypro.2016.10.107}
  {\path{doi:10.1016/j.egypro.2016.10.107}}.
\newline\urlprefix\url{https://doi.org/10.1016/j.egypro.2016.10.107}

\bibitem{GTM2018}
J.~Deign,
  \href{https://www.greentechmedia.com/articles/read/record-low-solar-plus-storage-price-in-xcel-solicitation}{{Xcel
  Attracts `Unprecedented' Low Prices for Solar and Wind Paired With Storage}}
  (2018).
\newline\urlprefix\url{https://www.greentechmedia.com/articles/read/record-low-solar-plus-storage-price-in-xcel-solicitation}

\bibitem{luo2015}
X.~Luo, J.~Wang, M.~Dooner, J.~Clarke,
  \href{https://doi.org/10.1016/j.apenergy.2014.09.081}{Overview of current
  development in electrical energy storage technologies and the application
  potential in power system operation}, Applied Energy 137 (2015) 511 -- 536.
\newblock \href {http://dx.doi.org/10.1016/j.apenergy.2014.09.081}
  {\path{doi:10.1016/j.apenergy.2014.09.081}}.
\newline\urlprefix\url{https://doi.org/10.1016/j.apenergy.2014.09.081}

\bibitem{gotz2016}
M.~Götz, J.~Lefebvre, F.~Mörs, A.~M. Koch, F.~Graf, S.~Bajohr, R.~Reimert,
  T.~Kolb, \href{https://doi.org/10.1016/j.renene.2015.07.066}{{Renewable
  Power-to-Gas: A technological and economic review}}, Renewable Energy 85
  (2016) 1371 -- 1390.
\newblock \href {http://dx.doi.org/10.1016/j.renene.2015.07.066}
  {\path{doi:10.1016/j.renene.2015.07.066}}.
\newline\urlprefix\url{https://doi.org/10.1016/j.renene.2015.07.066}

\bibitem{RONSCH2016276}
S.~Rönsch, J.~Schneider, S.~Matthischke, M.~Schlüter, M.~Götz, J.~Lefebvre,
  P.~Prabhakaran, S.~Bajohr,
  \href{https://doi.org/10.1016/j.fuel.2015.10.111}{{Review on methanation –
  From fundamentals to current projects}}, Fuel 166 (2016) 276 -- 296.
\newblock \href {http://dx.doi.org/10.1016/j.fuel.2015.10.111}
  {\path{doi:10.1016/j.fuel.2015.10.111}}.
\newline\urlprefix\url{https://doi.org/10.1016/j.fuel.2015.10.111}

\bibitem{FuelCell16}
{The Fuel Cell Industry Review 2016}, Tech. rep., {E4tech} (2016).

\bibitem{HELMETH}
\href{http://www.helmeth.eu/images/joomlaplates/documents/PI_2018_009_Power\%20to\%20Gas\%20with\%20High\%20Efficiency.pdf}{{Power-to-Gas
  with High Efficiency}} (2018).
\newline\urlprefix\url{http://www.helmeth.eu/images/joomlaplates/documents/PI_2018_009_Power\%20to\%20Gas\%20with\%20High\%20Efficiency.pdf}

\bibitem{IJSEPM1574}
H.~Lund, P.~Østergaard, D.~Connolly, I.~Ridjan, B.~Mathiesen, F.~Hvelplund,
  J.~Thellufsen, P.~Sorknæs,
  \href{https://doi.org/10.5278/ijsepm.2016.11.2}{{Energy Storage and Smart
  Energy Systems}}, International Journal of Sustainable Energy Planning and
  Management 11~(0).
\newblock \href {http://dx.doi.org/10.5278/ijsepm.2016.11.2}
  {\path{doi:10.5278/ijsepm.2016.11.2}}.
\newline\urlprefix\url{https://doi.org/10.5278/ijsepm.2016.11.2}

\bibitem{4957254}
B.~V. Mathiesen, H.~Lund,
  \href{https://doi.org/10.1049/iet-rpg:20080049}{Comparative analyses of seven
  technologies to facilitate the integration of fluctuating renewable energy
  sources}, IET Renewable Power Generation 3~(2) (2009) 190--204.
\newblock \href {http://dx.doi.org/10.1049/iet-rpg:20080049}
  {\path{doi:10.1049/iet-rpg:20080049}}.
\newline\urlprefix\url{https://doi.org/10.1049/iet-rpg:20080049}

\bibitem{gie}
G.~I. Europe, Gas storage data, \url{https://agsi.gie.eu/}.

\bibitem{GCBB:GCBB12205}
F.~Creutzig, N.~H. Ravindranath, G.~Berndes, S.~Bolwig, R.~Bright,
  F.~Cherubini, H.~Chum, E.~Corbera, M.~Delucchi, A.~Faaij, J.~Fargione,
  H.~Haberl, G.~Heath, O.~Lucon, R.~Plevin, A.~Popp, C.~Robledo-Abad, S.~Rose,
  P.~Smith, A.~Stromman, S.~Suh, O.~Masera,
  \href{http://dx.doi.org/10.1111/gcbb.12205}{Bioenergy and climate change
  mitigation: an assessment}, GCB Bioenergy 7~(5) (2015) 916--944.
\newblock \href {http://dx.doi.org/10.1111/gcbb.12205}
  {\path{doi:10.1111/gcbb.12205}}.
\newline\urlprefix\url{http://dx.doi.org/10.1111/gcbb.12205}

\bibitem{LINDFELDT20101836}
E.~G. Lindfeldt, M.~Saxe, M.~Magnusson, F.~Mohseni, {Strategies for a road
  transport system based on renewable resources – The case of an
  import-independent Sweden in 2025}, Applied Energy 87~(6) (2010) 1836 --
  1845.
\newblock \href {http://dx.doi.org/10.1016/j.apenergy.2010.02.011}
  {\path{doi:10.1016/j.apenergy.2010.02.011}}.

\bibitem{LUND2015389}
R.~Lund, B.~V. Mathiesen,
  \href{https://doi.org/10.1016/j.apenergy.2015.01.013}{Large combined heat and
  power plants in sustainable energy systems}, Applied Energy 142~(Supplement
  C) (2015) 389 -- 395.
\newblock \href {http://dx.doi.org/10.1016/j.apenergy.2015.01.013}
  {\path{doi:10.1016/j.apenergy.2015.01.013}}.
\newline\urlprefix\url{https://doi.org/10.1016/j.apenergy.2015.01.013}

\bibitem{Pires2011}
J.~Pires, F.~Martins, M.~Alvim-Ferraz, M.~Simões,
  \href{https://doi.org/10.1016/j.cherd.2011.01.028}{Recent developments on
  carbon capture and storage: An overview}, Chemical Engineering Research and
  Design 89~(9) (2011) 1446 -- 1460, special Issue on Carbon Capture \&
  Storage.
\newblock \href {http://dx.doi.org/10.1016/j.cherd.2011.01.028}
  {\path{doi:10.1016/j.cherd.2011.01.028}}.
\newline\urlprefix\url{https://doi.org/10.1016/j.cherd.2011.01.028}

\bibitem{Boots2014}
M.~E. Boot-Handford, J.~C. Abanades, E.~J. Anthony, M.~J. Blunt, S.~Brandani,
  N.~Mac~Dowell, J.~R. Fernandez, M.-C. Ferrari, R.~Gross, J.~P. Hallett, R.~S.
  Haszeldine, P.~Heptonstall, A.~Lyngfelt, Z.~Makuch, E.~Mangano, R.~T.~J.
  Porter, M.~Pourkashanian, G.~T. Rochelle, N.~Shah, J.~G. Yao, P.~S. Fennell,
  \href{http://dx.doi.org/10.1039/C3EE42350F}{Carbon capture and storage
  update}, Energy Environ. Sci. 7 (2014) 130--189.
\newblock \href {http://dx.doi.org/10.1039/C3EE42350F}
  {\path{doi:10.1039/C3EE42350F}}.
\newline\urlprefix\url{http://dx.doi.org/10.1039/C3EE42350F}

\bibitem{LEUNG2014426}
D.~Y. Leung, G.~Caramanna, M.~M. Maroto-Valer,
  \href{https://doi.org/10.1016/j.rser.2014.07.093}{An overview of current
  status of carbon dioxide capture and storage technologies}, Renewable and
  Sustainable Energy Reviews 39 (2014) 426 -- 443.
\newblock \href {http://dx.doi.org/10.1016/j.rser.2014.07.093}
  {\path{doi:10.1016/j.rser.2014.07.093}}.
\newline\urlprefix\url{https://doi.org/10.1016/j.rser.2014.07.093}

\bibitem{McLaren2012}
D.~McLaren, \href{https://doi.org/10.1016/j.psep.2012.10.005}{A comparative
  global assessment of potential negative emissions technologies}, Process
  Safety and Environmental Protection 90~(6) (2012) 489 -- 500, special Issue:
  Negative emissions technology.
\newblock \href {http://dx.doi.org/10.1016/j.psep.2012.10.005}
  {\path{doi:10.1016/j.psep.2012.10.005}}.
\newline\urlprefix\url{https://doi.org/10.1016/j.psep.2012.10.005}

\bibitem{LORANGESEIGO2014848}
S.~L. Seigo, S.~Dohle, M.~Siegrist,
  \href{https://doi.org/10.1016/j.rser.2014.07.017}{Public perception of carbon
  capture and storage (ccs): A review}, Renewable and Sustainable Energy
  Reviews 38 (2014) 848 -- 863.
\newblock \href {http://dx.doi.org/10.1016/j.rser.2014.07.017}
  {\path{doi:10.1016/j.rser.2014.07.017}}.
\newline\urlprefix\url{https://doi.org/10.1016/j.rser.2014.07.017}

\bibitem{LUND2012469}
H.~Lund, B.~V. Mathiesen,
  \href{https://doi.org/10.1016/j.energy.2012.06.002}{The role of carbon
  capture and storage in a future sustainable energy system}, Energy 44~(1)
  (2012) 469 -- 476, integration and Energy System Engineering, European
  Symposium on Computer-Aided Process Engineering 2011.
\newblock \href {http://dx.doi.org/10.1016/j.energy.2012.06.002}
  {\path{doi:10.1016/j.energy.2012.06.002}}.
\newline\urlprefix\url{https://doi.org/10.1016/j.energy.2012.06.002}

\bibitem{FASIHI2016243}
M.~Fasihi, D.~Bogdanov, C.~Breyer,
  \href{https://doi.org/10.1016/j.egypro.2016.10.115}{{Techno-Economic
  Assessment of Power-to-Liquids (PtL) Fuels Production and Global Trading
  Based on Hybrid PV-Wind Power Plants}}, Energy Procedia 99 (2016) 243 -- 268,
  10th International Renewable Energy Storage Conference, IRES 2016, 15-17
  March 2016, Düsseldorf, Germany.
\newblock \href {http://dx.doi.org/10.1016/j.egypro.2016.10.115}
  {\path{doi:10.1016/j.egypro.2016.10.115}}.
\newline\urlprefix\url{https://doi.org/10.1016/j.egypro.2016.10.115}

\bibitem{su9020306}
M.~Fasihi, D.~Bogdanov, C.~Breyer,
  \href{https://doi.org/10.3390/su9020306}{Long-term hydrocarbon trade options
  for the maghreb region and europe—renewable energy based synthetic fuels
  for a net zero emissions world}, Sustainability 9~(2).
\newblock \href {http://dx.doi.org/10.3390/su9020306}
  {\path{doi:10.3390/su9020306}}.
\newline\urlprefix\url{https://doi.org/10.3390/su9020306}

\bibitem{Keith1654}
D.~W. Keith, \href{https://doi.org/10.1126/science.1175680}{Why capture co2
  from the atmosphere?}, Science 325~(5948) (2009) 1654--1655.
\newblock \href {http://dx.doi.org/10.1126/science.1175680}
  {\path{doi:10.1126/science.1175680}}.
\newline\urlprefix\url{https://doi.org/10.1126/science.1175680}

\bibitem{Williamson2016}
P.~Williamson, \href{https://doi.org/10.1038/530153a}{Emissions reduction:
  Scrutinize co2 removal methods}, Nature 530 (2016) 153--155.
\newblock \href {http://dx.doi.org/10.1038/530153a}
  {\path{doi:10.1038/530153a}}.
\newline\urlprefix\url{https://doi.org/10.1038/530153a}

\bibitem{Fuss2014}
S.~Fuss, J.~G. Canadell, G.~P. Peters, M.~Tavoni, R.~M. Andrew, P.~Ciais, R.~B.
  Jackson, C.~D. Jones, F.~Kraxner, N.~Nakicenovic, C.~Le~Qu{\'e}r{\'e}, M.~R.
  Raupach, A.~Sharifi, P.~Smith, Y.~Yamagata,
  \href{http://dx.doi.org/10.1038/nclimate2392}{Betting on negative emissions},
  Nature Climate Change 4 (2014) 850 EP --.
\newblock \href {http://dx.doi.org/10.1038/nclimate2392}
  {\path{doi:10.1038/nclimate2392}}.
\newline\urlprefix\url{http://dx.doi.org/10.1038/nclimate2392}

\bibitem{Schleussner2016}
C.-F. Schleussner, J.~Rogelj, M.~Schaeffer, T.~Lissner, R.~Licker, E.~M.
  Fischer, R.~Knutti, A.~Levermann, K.~Frieler, W.~Hare,
  \href{http://dx.doi.org/10.1038/nclimate3096}{{Science and policy
  characteristics of the Paris Agreement temperature goal}}, Nature Climate
  Change 6 (2016) 827 EP --, perspective.
\newblock \href {http://dx.doi.org/10.1038/nclimate3096}
  {\path{doi:10.1038/nclimate3096}}.
\newline\urlprefix\url{http://dx.doi.org/10.1038/nclimate3096}

\bibitem{vanVuuren2017}
D.~P. van Vuuren, A.~F. Hof, M.~A.~E. van Sluisveld, K.~Riahi,
  \href{https://doi.org/10.1038/s41560-017-0055-2}{Open discussion of negative
  emissions is urgently needed}, Nature Energy 2~(12) (2017) 902--904.
\newblock \href {http://dx.doi.org/10.1038/s41560-017-0055-2}
  {\path{doi:10.1038/s41560-017-0055-2}}.
\newline\urlprefix\url{https://doi.org/10.1038/s41560-017-0055-2}

\bibitem{Rockstrom1269}
J.~Rockstr{\"o}m, O.~Gaffney, J.~Rogelj, M.~Meinshausen, N.~Nakicenovic, H.~J.
  Schellnhuber, \href{https://doi.org/10.1126/science.aah3443}{A roadmap for
  rapid decarbonization}, Science 355~(6331) (2017) 1269--1271.
\newblock \href {http://dx.doi.org/10.1126/science.aah3443}
  {\path{doi:10.1126/science.aah3443}}.
\newline\urlprefix\url{https://doi.org/10.1126/science.aah3443}

\bibitem{Smith2015}
P.~Smith, S.~J. Davis, F.~Creutzig, S.~Fuss, J.~Minx, B.~Gabrielle, E.~Kato,
  R.~B. Jackson, A.~Cowie, E.~Kriegler, D.~P. van Vuuren, J.~Rogelj, P.~Ciais,
  J.~Milne, J.~G. Canadell, D.~McCollum, G.~Peters, R.~Andrew, V.~Krey,
  G.~Shrestha, P.~Friedlingstein, T.~Gasser, A.~Gr{\"u}bler, W.~K. Heidug,
  M.~Jonas, C.~D. Jones, F.~Kraxner, E.~Littleton, J.~Lowe, J.~R. Moreira,
  N.~Nakicenovic, M.~Obersteiner, A.~Patwardhan, M.~Rogner, E.~Rubin,
  A.~Sharifi, A.~Torvanger, Y.~Yamagata, J.~Edmonds, C.~Yongsung,
  \href{http://dx.doi.org/10.1038/nclimate2870}{{Biophysical and economic
  limits to negative CO2 emissions}}, Nature Climate Change 6, review Article.
\newblock \href {http://dx.doi.org/10.1038/nclimate2870}
  {\path{doi:10.1038/nclimate2870}}.
\newline\urlprefix\url{http://dx.doi.org/10.1038/nclimate2870}

\bibitem{Anderson2016}
K.~Anderson, G.~Peters, \href{https://doi.org/10.1126/science.aah4567}{The
  trouble with negative emissions}, Science 354~(6309) (2016) 182--183.
\newblock \href
  {http://arxiv.org/abs/http://science.sciencemag.org/content/354/6309/182.full.pdf}
  {\path{arXiv:http://science.sciencemag.org/content/354/6309/182.full.pdf}},
  \href {http://dx.doi.org/10.1126/science.aah4567}
  {\path{doi:10.1126/science.aah4567}}.
\newline\urlprefix\url{https://doi.org/10.1126/science.aah4567}

\bibitem{Vaughan2016}
N.~E. Vaughan, C.~Gough,
  \href{http://dx.doi.org/10.1088/1748-9326/11/9/095003}{Expert assessment
  concludes negative emissions scenarios may not deliver}, Environmental
  Research Letters 11~(9) (2016) 095003.
\newblock \href {http://dx.doi.org/10.1088/1748-9326/11/9/095003}
  {\path{doi:10.1088/1748-9326/11/9/095003}}.
\newline\urlprefix\url{http://dx.doi.org/10.1088/1748-9326/11/9/095003}

\bibitem{EUsurvey2017}
\href{https://doi.org/10.2834/92702}{{Special Eurobarometer 459 Report: Climate
  Change}} (2017).
\newblock \href {http://dx.doi.org/10.2834/92702} {\path{doi:10.2834/92702}}.
\newline\urlprefix\url{https://doi.org/10.2834/92702}

\bibitem{Orsted2017}
\href{https://orsted.com/en/Barometer}{{Ørsted Green Energy Barometer}}
  (2017).
\newline\urlprefix\url{https://orsted.com/en/Barometer}

\bibitem{AEE2016}
\href{https://www.unendlich-viel-energie.de/media/file/426.AEE_RK29_Internationale_Akzeptanzumfragen_Mrz16.pdf}{{Die
  Akzeptanz für Erneuerbaren Energien im Spiegel von Umfragen in
  Industriestaaten}} (2016).
\newline\urlprefix\url{https://www.unendlich-viel-energie.de/media/file/426.AEE_RK29_Internationale_Akzeptanzumfragen_Mrz16.pdf}

\bibitem{MARUYAMA20072761}
Y.~Maruyama, M.~Nishikido, T.~Iida,
  \href{https://doi.org/10.1016/j.enpol.2006.12.010}{{The rise of community
  wind power in Japan: Enhanced acceptance through social innovation}}, Energy
  Policy 35~(5) (2007) 2761 -- 2769.
\newblock \href {http://dx.doi.org/10.1016/j.enpol.2006.12.010}
  {\path{doi:10.1016/j.enpol.2006.12.010}}.
\newline\urlprefix\url{https://doi.org/10.1016/j.enpol.2006.12.010}

\bibitem{JOBERT20072751}
A.~Jobert, P.~Laborgne, S.~Mimler,
  \href{https://doi.org/10.1016/j.enpol.2006.12.005}{{Local acceptance of wind
  energy: Factors of success identified in French and German case studies}},
  Energy Policy 35~(5) (2007) 2751 -- 2760.
\newblock \href {http://dx.doi.org/10.1016/j.enpol.2006.12.005}
  {\path{doi:10.1016/j.enpol.2006.12.005}}.
\newline\urlprefix\url{https://doi.org/10.1016/j.enpol.2006.12.005}

\bibitem{ENEVOLDSEN2016178}
P.~Enevoldsen, B.~K. Sovacool,
  \href{https://doi.org/10.1016/j.rser.2015.08.041}{{Examining the social
  acceptance of wind energy: Practical guidelines for onshore wind project
  development in France}}, Renewable and Sustainable Energy Reviews 53 (2016)
  178 -- 184.
\newblock \href {http://dx.doi.org/10.1016/j.rser.2015.08.041}
  {\path{doi:10.1016/j.rser.2015.08.041}}.
\newline\urlprefix\url{https://doi.org/10.1016/j.rser.2015.08.041}

\bibitem{Schlachtberger2018}
D.~Schlachtberger, T.~Brown, M.~Schäfer, S.~Schramm, M.~Greiner,
  \href{https://arxiv.org/abs/1803.09711}{{Cost optimal scenarios of a future
  highly renewable European electricity system: Exploring the influence of
  weather data, cost parameters and policy constraints}}.
\newline\urlprefix\url{https://arxiv.org/abs/1803.09711}

\bibitem{COHEN20144}
J.~J. Cohen, J.~Reichl, M.~Schmidthaler,
  \href{https://doi.org/10.1016/j.energy.2013.12.056}{Re-focussing research
  efforts on the public acceptance of energy infrastructure: A critical
  review}, Energy 76 (2014) 4 -- 9.
\newblock \href {http://dx.doi.org/10.1016/j.energy.2013.12.056}
  {\path{doi:10.1016/j.energy.2013.12.056}}.
\newline\urlprefix\url{https://doi.org/10.1016/j.energy.2013.12.056}

\bibitem{GALVIN2018114}
R.~Galvin, \href{https://doi.org/10.1016/j.erss.2018.01.022}{{Trouble at the
  end of the line: Local activism and social acceptance in low-carbon
  electricity transmission in Lower Franconia, Germany}}, Energy Research \&
  Social Science 38 (2018) 114 -- 126.
\newblock \href {http://dx.doi.org/10.1016/j.erss.2018.01.022}
  {\path{doi:10.1016/j.erss.2018.01.022}}.
\newline\urlprefix\url{https://doi.org/10.1016/j.erss.2018.01.022}

\bibitem{PFENNINGER2017211}
S.~Pfenninger, J.~DeCarolis, L.~Hirth, S.~Quoilin, I.~Staffell,
  \href{https://doi.org/10.1016/j.enpol.2016.11.046}{{The importance of open
  data and software: Is energy research lagging behind?}}, Energy Policy 101
  (2017) 211 -- 215.
\newblock \href {http://dx.doi.org/10.1016/j.enpol.2016.11.046}
  {\path{doi:10.1016/j.enpol.2016.11.046}}.
\newline\urlprefix\url{https://doi.org/10.1016/j.enpol.2016.11.046}

\bibitem{PFENNINGER201863}
S.~Pfenninger, L.~Hirth, I.~Schlecht, E.~Schmid, F.~Wiese, T.~Brown, C.~Davis,
  M.~Gidden, H.~Heinrichs, C.~Heuberger, S.~Hilpert, U.~Krien, C.~Matke,
  A.~Nebel, R.~Morrison, B.~Müller, G.~Pleßmann, M.~Reeg, J.~C. Richstein,
  A.~Shivakumar, I.~Staffell, T.~Tröndle, C.~Wingenbach,
  \href{https://doi.org/10.1016/j.esr.2017.12.002}{{Opening the black box of
  energy modelling: Strategies and lessons learned}}, Energy Strategy Reviews
  19 (2018) 63 -- 71.
\newblock \href {http://dx.doi.org/10.1016/j.esr.2017.12.002}
  {\path{doi:10.1016/j.esr.2017.12.002}}.
\newline\urlprefix\url{https://doi.org/10.1016/j.esr.2017.12.002}

\bibitem{Child2018321}
M.~Child, O.~Koskinen, L.~Linnanen, C.~Breyer,
  \href{https://doi.org/10.1016/j.rser.2018.03.079}{Sustainability guardrails
  for energy scenarios of the global energy transition}, Renewable and
  Sustainable Energy Reviews 91 (2018) 321 -- 334.
\newblock \href {http://dx.doi.org/10.1016/j.rser.2018.03.079}
  {\path{doi:10.1016/j.rser.2018.03.079}}.
\newline\urlprefix\url{https://doi.org/10.1016/j.rser.2018.03.079}

\bibitem{IRENA2018}
\href{http://www.irena.org/-/media/Files/IRENA/Agency/Publication/2018/Jan/IRENA_2017_Power_Costs_2018.pdf}{{Renewable
  Power Generation Costs in 2017}}, Tech. rep., International Renewable Energy
  Agency (IRENA) (2018).
\newline\urlprefix\url{http://www.irena.org/-/media/Files/IRENA/Agency/Publication/2018/Jan/IRENA_2017_Power_Costs_2018.pdf}

\bibitem{Joskow2011}
P.~L. Joskow, \href{https://doi.org/10.1257/aer.101.3.238}{Comparing the costs
  of intermittent and dispatchable electricity generating technologies}, The
  American Economic Review 101~(3) (2011) 238--241.
\newblock \href {http://dx.doi.org/10.1257/aer.101.3.238}
  {\path{doi:10.1257/aer.101.3.238}}.
\newline\urlprefix\url{https://doi.org/10.1257/aer.101.3.238}

\bibitem{B809990C}
M.~Z. Jacobson, \href{http://dx.doi.org/10.1039/B809990C}{Review of solutions
  to global warming{,} air pollution{,} and energy security}, Energy Environ.
  Sci. 2 (2009) 148--173.
\newblock \href {http://dx.doi.org/10.1039/B809990C}
  {\path{doi:10.1039/B809990C}}.
\newline\urlprefix\url{http://dx.doi.org/10.1039/B809990C}

\bibitem{VERBRUGGEN201416}
A.~Verbruggen, E.~Laes, S.~Lemmens,
  \href{https://doi.org/10.1016/j.rser.2014.01.008}{Assessment of the actual
  sustainability of nuclear fission power}, Renewable and Sustainable Energy
  Reviews 32 (2014) 16 -- 28.
\newblock \href {http://dx.doi.org/10.1016/j.rser.2014.01.008}
  {\path{doi:10.1016/j.rser.2014.01.008}}.
\newline\urlprefix\url{https://doi.org/10.1016/j.rser.2014.01.008}

\bibitem{sma2012}
\href{https://www.sma.de/en/newsroom/current-news/news-details/news/3943-tokelau-becomes-the-worlds-first-100-solar-powered-country.html}{Tokelau
  becomes the world's first 100\% solar powered country}, press Release (Nov
  2012).
\newblock \href
  {http://dx.doi.org/http://dx.doi.org/10.1016/j.egypro.2016.10.107}
  {\path{doi:http://dx.doi.org/10.1016/j.egypro.2016.10.107}}.
\newline\urlprefix\url{https://www.sma.de/en/newsroom/current-news/news-details/news/3943-tokelau-becomes-the-worlds-first-100-solar-powered-country.html}

\bibitem{AEMO2017}
\href{https://www.aemo.com.au/Media-Centre/AEMO-publishes-final-report-into-the-South-Australian-state-wide-power-outage}{{Black
  System South Australia 28 September 2016 - Final Report}}, Tech. rep.,
  Australian Energy Market Operator (AEMO) (March 2017).
\newline\urlprefix\url{https://www.aemo.com.au/Media-Centre/AEMO-publishes-final-report-into-the-South-Australian-state-wide-power-outage}

\end{thebibliography}

\end{document}